\documentclass[AMA,STIX1COL]{WileyNJD-v2}
\usepackage{moreverb}

\usepackage{makecell}
\usepackage{enumerate}
\usepackage{tabularx}
\usepackage{graphics}
\usepackage{graphicx}
\usepackage{subfig}
\usepackage{dsfont}
\usepackage{blkarray}
\usepackage{amsmath}
\soulregister\cite7

\newcommand\BibTeX{{\rmfamily B\kern-.05em \textsc{i\kern-.025em b}\kern-.08em
T\kern-.1667em\lower.7ex\hbox{E}\kern-.125emX}}

\articletype{Article Type}%

\received{<day> <Month>, <year>}
\revised{<day> <Month>, <year>}
\accepted{<day> <Month>, <year>}

\usepackage{framed,color}

\definecolor{shadecolor}{rgb}{0.9,0.9,0.9}

\newcommand{\comments}[1]{}

\def\P{P}					

\newcommand{\beq}{\begin{equation}}
\newcommand{\eeq}{\end{equation}}
\newcommand{\bal}{\begin{aligned}}
\newcommand{\eal}{\end{aligned}}

\newcommand{\be}{\begin{equation}}
\newcommand{\ee}{\end{equation}}
\newcommand{\bd}{\begin{displaymath}}
\newcommand{\ed}{\end{displaymath}}
\newcommand{\BE}{\begin{eqnarray}}
\newcommand{\EE}{\end{eqnarray}}

\begin{document}

\title{Network meta-analysis and random walks}

\author[1]{Annabel L. Davies}
\author[2]{Theodoros Papakonstantinou}
\author[2]{Adriani Nikolakopoulou}
\author[2]{Gerta R{\"u}cker}
\author[1,3]{Tobias Galla}

\authormark{DAVIES et al}

\address[1]{\orgdiv{Theoretical Physics, Department of Physics and Astronomy, School of Natural Sciences}, \orgname{The University of Manchester}, \orgaddress{\state{Manchester}, \country{UK}}}

\address[2]{\orgdiv{Institute of Medical Biometry and Statistics, Faculty of Medicine and Medical Centre},\orgname{University of Freiburg}, \orgaddress{\state{Freiburg}, \country{Germany}}}

\address[3]{\orgdiv{Instituto de F\'isica Interdisciplinar y Sistemas Complejos}, \orgname{IFISC (CSIC-UIB), Campus Universitat Illes Balears}, \orgaddress{\state{Palma de Mallorca}, \country{Spain}}}

\corres{Annabel L. Davies, Theoretical Physics, Department of Physics and Astronomy, School of Natural Sciences, The University of Manchester, Manchester M13 9PL, United Kingdom. \email{annabel.davies@postgrad.manchester.ac.uk}}

\jnlcitation{\cname{%
\author{Davies AL} et al.} (\cyear{2021}), 
\ctitle{Network meta-analysis and random walks}}

\abstract[Abstract]{ 
Network meta-analysis (NMA) is a central tool for evidence synthesis in clinical research. The results of an NMA depend critically on the quality of evidence being pooled. In assessing the validity of an NMA, it is therefore important to know the proportion contributions of each direct treatment comparison to each network treatment effect. The construction of proportion contributions is based on the observation that each row of the hat matrix represents a so-called `evidence flow network' for each treatment comparison. However, the existing algorithm used to calculate these values is associated with ambiguity according to the selection of paths. In this work we present a novel analogy between NMA and random walks. We use this analogy to derive closed-form expressions for the proportion contributions. A random walk on a graph is a stochastic process that describes a succession of random ‘hops’ between vertices which are connected by an edge. The weight of an edge relates to the probability that the walker moves along that edge. We use the graph representation of NMA to construct the transition matrix for a random walk on the network of evidence. We show that the net number of times a walker crosses each edge of the network is related to the evidence flow network. By then defining a random walk on the directed evidence flow network, we derive analytically the matrix of proportion contributions. The random-walk approach, in addition to being computationally more efficient, has none of the associated ambiguity of the existing algorithm. 
}

\maketitle

\section{Introduction}
\label{intro}

Network meta-analysis (NMA) has been established as a central tool of evidence synthesis in clinical research\cite{TSD2, DIAS:2018, SALANTI:2012}. Combining direct and indirect evidence from multiple randomised controlled trials, NMA makes it possible to compare interventions that have not been tested together in any trial \cite{Lu:Ades:2004, Hig:White:1996, Lumley:2002}. The term `network meta-analysis' derives from the fact that one can mathematically represent the collection of interventions and trials as a graph. A graph consists of a set of nodes and a set of edges connecting pairs of nodes. The nodes of an NMA graph represent the different treatment options, and edges are comparisons made between the treatments in the trials. In line with R{\"u}cker (2012)\cite{Rucker:2012} we will refer to networks of treatment options and comparisons between treatments as `meta-analytic graphs'.

An NMA combines data from multiple trials, each comparing different combinations of treatment options. The accuracy of the conclusions from an NMA depends on potential biases associated with individual trials, and on assumptions such as between-trial homogeneity and consistency between direct and indirect evidence. In this context it is useful to study the so-called `flow of evidence' \cite{konig:2013} in the network. This describes the influence different network components have on the estimates of treatment effects. For example, the comparison between two particular treatments may enter as indirect evidence into the estimate of the relative  effect of two different nodes in the network. 
Understanding how exactly evidence flows in the graph then allows one to assess the impact of potential bias originating from different pieces of evidence in the network \cite{konig:2013, Papakon:2018, Nikola:2020}.

Previous literature has, for example, looked at the relative influence of direct evidence compared to indirect evidence \cite{Lu:2011, Senn:2012}. Other work has been concerned with measures of network geometry, capturing the frequency with which different comparisons are represented in the trials underpinning an NMA \cite{Salanti:2008, Salanti:2008b}. One then asks how the network structure affects NMA estimates of treatment effects, heterogeneity and rank metrics\cite{Salanti:2008, Salanti:2008b, Davies:2020, Tonin:2019, Veroniki:2018, Kibret:2014}. K{\"{o}}nig et al (2013) \cite{konig:2013} observed that in a two-step (`aggregate') NMA model\cite{Lu:2011} each row of the `hat matrix' represents an evidence flow network for a particular treatment effect. K{\"{o}}nig et al then visualised the evidence flow on weighted directed acyclic graphs in which nodes represent treatments, and edges indicate the direction and quantity of evidence flow through each direct comparison. Based on this observation, Papakonstantinou et al (2018) \cite{Papakon:2018} introduced `streams' of evidence and developed a numerical algorithm to calculate these streams. The streams of evidence are then used to derive the `proportion contribution' of each direct comparison to each treatment effect in the graph. This allows one to quantify how limitations of individual studies impact on the estimates obtained from the network. Indeed, the algorithm in Papakonstantinou et al. is implemented in software such as CINeMA (confidence in network meta-analysis) \cite{Nikola:2020} and ROB-MEN (risk-of-bias due to missing evidence in NMA)\cite{Chiocchia:2021}, used in clinical practice for the evaluation of results from an NMA.

More widely, the study of networks plays a key role in a variety of disciplines including ecology, economics, electrical engineering and sociology\cite{Mendes:2003, Newman:2018, Estrada:2011}. Through the representation of treatment options and comparisons in trials as a graph, one can therefore take advantage of the extensive literature in network theory, and of ideas developed in the disciplines in which networks are studied. For example one of us \cite{Rucker:2012} used the graph representation of NMA to make the connection between meta-analytic and electrical networks. This allows one to demonstrate that graph theoretical tools routinely applied to electrical networks are also of use in NMA. This approach has since led to advancements in NMA methodology such as frequentist ranking methods \cite{Rucker:2015} and component NMA \cite{Rucker:2020}. It is also the basis for the software package {\it netmeta} \cite{netmeta:2021}.

In this paper we present a new analogy between random walks and NMA. A random walk on a graph is a stochastic process consisting of a succession of ‘hops’ between vertices connected by edges. Random walks are of interest for a wide range of applications, including statistical physics, biology, ecology, genetics, transport and economics (for a selection of references see \cite{Codling,Okubo,Isichenko,Ewens,Mantegna}). Random walks are also a popular tool to study the properties of networks themselves \cite{Noh:2004, Lov:1994, MASUDA20171}.

It is well known that there is a connection between random walks and electrical networks  \cite{Kakutani:1945, Kemeny:1966, Kelly:1979, Doyle:2000}. In this context, edges of the electric network are conducting connections (wires). The correspondence between random walks and electrical networks can be established by asserting that the probability that a random walker currently at node $a$ moves to node $b$ in the next step is proportional to the conductance (inverse resistance) of the edge connecting $a$ and $b$. Quantities in the electrical network such as currents along edges or electric potentials at the nodes then have an interpretation in the random-walk picture. For further details we refer to \cite{Doyle:2000}.

Motivated by the connection between electrical networks and NMA on the one hand, and that of electrical networks and random walks on the other, we construct a random walk on the \textit{meta-analytic network}. We show that the random-walk picture we develop can be used to study the flow of evidence in the NMA network. In particular there is a random-walk interpretation of the elements in the hat matrix. Further, we construct a second random-walk model, this time on the \textit{evidence flow network}. From this we derive an analytical expression for proportion contributions which overcomes the limitations of Papakonstantinou et al \cite{Papakon:2018}. In particular, the algorithm in Papakonstantinou et al selects only a subset of paths on the evidence flow network. This means that paths of evidence that potentially contribute risk of bias are missed. Furthermore, the paths identified by the algorithm are not unambiguous and instead depend on the order in which certain steps are carried out. In contrast, the random-walk approach identifies all possible paths of evidence. The method delivers an unambiguous analytical result for proportion contributions, and less computational effort is required than for the numerical algorithm. In addition, unlike the method in Papakonstantinou et al, the random-walk approach is able to handle networks with multi-arm trials.

The remainder of this paper is set out as follows:   We present a motivating data set in Section \ref{real_data}. In Section \ref{NMA_model}  we provide the relevant background information. We describe an aggregate-level frequentist NMA model and show how the associated hat matrix can be interpreted as evidence flow. In Section \ref{analogies} we introduce the analogies between NMA, electrical networks and random walks. Using the analogies to electrical networks in both the NMA and random-walk literature, we then express the flow of evidence in an NMA in terms of properties of random walks on the aggregate network. In Section \ref{prop} we introduce a second random-walk model, now on the directed evidence flow network. We use this to analytically derive the matrix of proportion contributions. In Section \ref{apply}, we apply our method to the motivating data set and demonstrate that the random-walk approach overcomes the limitations of the numerical algorithm previously proposed by Papakonstantinou et al (2018) \cite{Papakon:2018}. We summarise our results in Section \ref{discuss} and discuss potential future impact of the analogy between NMA and random walks.

\section{Motivating Example}
\label{real_data}
\begin{figure}
    \centering
    \includegraphics[width=0.8\linewidth]{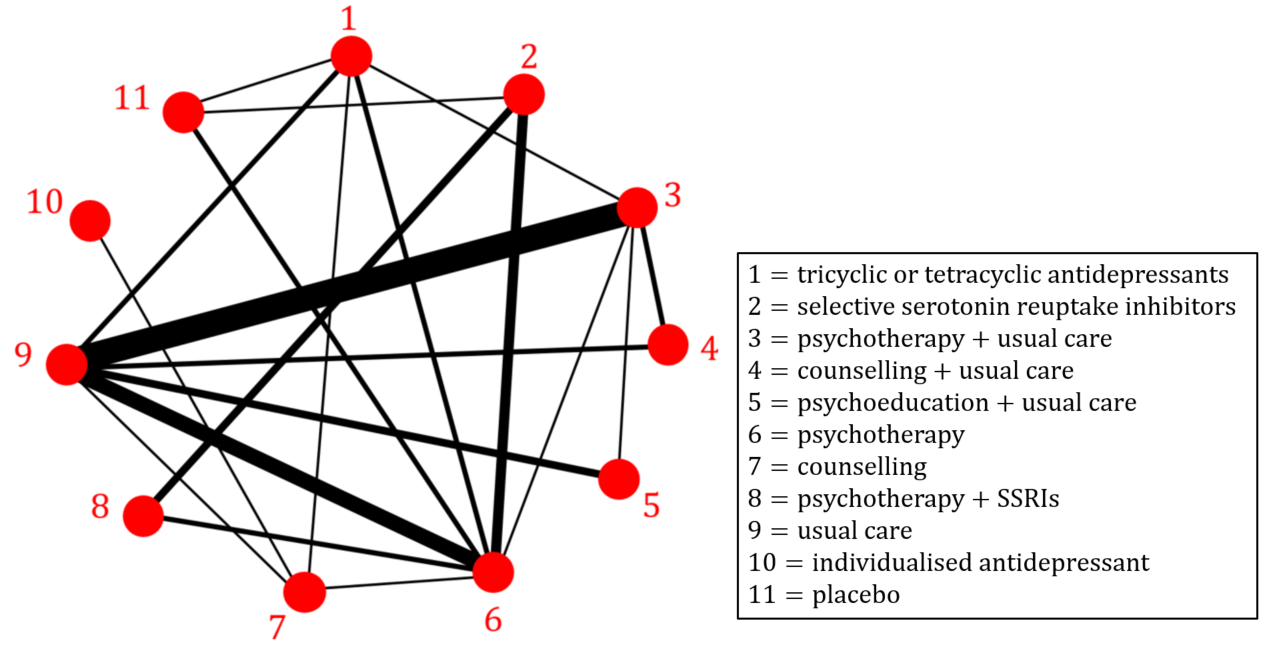}
    \caption{~A network of psychological treatments for depression (original data from Linde et al (2013) \cite{Linde:2013}; presented in R{\"u}cker and Schwarzer (2014) \cite{Rucker:2014}). We use numerical labels from $1$ to $11$, these are the same as in \cite{Rucker:2014}. Two treatments are connected by an edge if a direct comparison of the two treatments was made in at least one trial; the edge width indicates the number of trials that make the comparison. The network contains one $4$-arm trial (comparing treatments 1-6-7-9), eight $3$-arm trials (3-5-9, 2-6-8, 1-6-11, 1-3-9, 2-6-11, 2-6-8, 3-6-9, and 3-4-9) and $17$ $2$-arm trials. Multi-arm trials are not explicitly indicated in the network graph. The data, including the number of trials per comparison, is described in detail in R{\"u}cker and Schwarzer (2014) \cite{Rucker:2014}.  }
    \label{fig:realdata}
\end{figure}

We use an NMA of psychological treatments for patients with depressive disorders \cite{Linde:2013} to motivate our work. The data is described in detail in 
R{\"u}cker and Schwarzer (2014) \cite{Rucker:2014}. For convenience we will occasionally refer to this as the `depression data set'. The NMA compares $N=11$ treatments based on $M=26$ randomised controlled trials. Of these one is a four-arm trial, eight are three-arm trials and 17 contain just two arms. In total, the trials provide $K=20$ pairs of treatments which are directly compared in at least one trial. The primary outcome of the trials was a binary variable representing patient response after treatment completion. The odds ratio (OR) was used as the measure of relative treatment effect. The graph representing this set of treatments and trials is shown in Figure \ref{fig:realdata}. Vertices in the graph are treatments, and edges represent comparisons between pairs of treatments (two vertices are connected if they have been directly compared in at least one trial). The graph therefore has $N=11$ vertices and $K=20$ edges. The thickness of the edges in the figure represent the number of trials making the different comparisons.

NMA aims at estimating treatment effects for all pairs of interventions within this network. One aim of our paper is to determine the contribution (as a proportion) of each direct comparison to these estimates.

\section{Network meta-analysis model}
\label{NMA_model}
\subsection{Definitions and notation}
Among the multiple equivalent frequentist formulations of NMA \cite{Lumley:2002, Lu:2011, Salanti:2008, Rucker:2014,  Efthimiou:2016} we choose a so-called `aggregate level' (or two-step) approach\cite{Lu:2011} to the graph theoretical model developed in R\" ucker (2012) \cite{Rucker:2012}.  R\"ucker's original (one-step) model is implemented in the R package {\it netmeta} \cite{netmeta:2021}. In Appendix \ref{App:Freq} we outline how the aggregate-level graph theoretical approach relates to other frequentist NMA models.

We consider a network of $N$ treatments, denoted $a=1,\dots,N$, and $M$ studies, $i=1,\dots,M$. Throughout this article we will use the lower case letters $a,b,c$ and $d$ to refer to treatment nodes. Occasionally we also use $x$ and $y$ as dummy indices referring to nodes in sums or products. Study $i$ compares a subset of $n_i$ treatments (i.e., $n_i$ is the number of treatments in trial $i$). We use a random-effects model where we focus on relative, rather than absolute effects. To this end, we write $Y_{i,ab}$ for the observed effect of treatment $b$ in trial $i$ relative to treatment $a$. We denote the variance associated with this observation by $\sigma_{i,ab}^2$. The heterogeneity, $\tau^2$, in the network can be estimated, for example, using the method-of-moments approach \cite{Jackson:2012}. The estimated heterogeneity is added to the within-trial variance estimate from each study to make the total variance $\sigma_{i,ab}^2 + \tau^2$.

Trial $i$ has $n_i$ arms and contributes $q_i = n_i(n_i-1)/2$ observed relative treatment effects and associated variances. For a trial with $n_i=2$,  comparing treatments $a$ and $b$, the weight assigned is given by the inverse variance, $w_{i,ab}=1/(\sigma_{i,ab}^2 + \tau^2)$. In order to account for correlations induced by multi-arm trials ($n_i\geq3$), we use an adjustment method described in detail in References \cite{Rucker:2012, Rucker:2014, Gutman:2004}. The method involves adjusting the variances associated with each pairwise comparison in a multi-arm trial. For multi-arm trial $i$ this results in $q_i\geq 3$ weights, $w_{i,ab}$, where $a$ and $b$ run through all treatments compared in that trial. This defines a complete sub-graph of $q_i$ two-arm trials which is equivalent to the multi-arm trial. 

\subsection{Aggregate-level description}\label{sec:aggregate}
The set of adjusted weights $\{w_{i,ab}\}$ for all trials $i=1,\hdots,M$ defines a network of $\sum_{i=1}^M q_i$ two-arm trials. This network is equivalent to the original network of $M$ (potentially multi-arm) trials in that the resulting relative treatment effect estimates from the network of two-arm trials described by $\{w_{i,ab}\}$ are the same as those obtained from the original network\cite{Rucker:2014}.

We write $M_{ab}$ for the set of trials $i\in\{1,\dots,M\}$ comparing treatments $a$ and $b$.  Using the weights $\{w_{i,ab}\}$, we perform a pairwise meta-analysis across each of the $K$ edges in the network. For the edge connecting nodes $a$ and $b$, the direct estimate is calculated as the weighted mean, 
\begin{align}
\label{eq:weight_mean}
    \hat\theta_{ab}^{\mathrm{dir}} = \frac{\sum_{i\in M_{ab}} w_{i,ab} Y_{i,ab} }{\sum_{i\in M_{ab}} w_{i,ab}}.
\end{align}
This results in $K$ \textit{direct} estimates of the relative treatment effects, $\hat\theta^{\mathrm{dir}}_{ab}$, which we collect in the vector $\boldsymbol{\hat{\theta}}^{\mathrm{dir}}$. The weight associated with the direct estimate $\hat{\theta}_{ab}^{\mathrm{dir}}$ (and to be used in the subsequent analysis) is given by 
\begin{align}
\label{eq:weight}
    w_{ab} = \sum_{i\in M_{ab}} w_{i,ab}.
\end{align}
The direct estimates of the relative treatment effects have been termed `aggregate' data \cite{konig:2013, Krahn:2013}. Therefore, Equations (\ref{eq:weight_mean}) and (\ref{eq:weight}) describe the observations and inverse-variance weights for an aggregate-level model.

\begin{figure}
    \centering
    \includegraphics[width=1\linewidth]{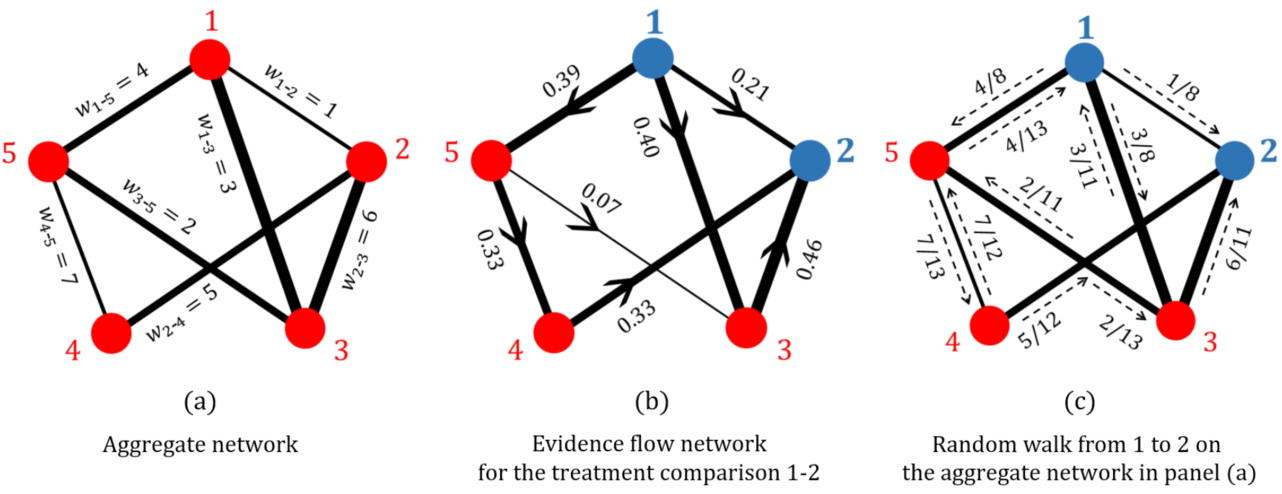}
    \caption{~ (a) A fictional example of an aggregate meta-analytic network with edges weighted and labelled by their respective (inverse-variance) weights. (b) The resulting evidence flow network for the comparison 1-2 from the aggregate network in (a); the comparison 1-2 is indicated by depicting these nodes and their labels in blue. Edges are directed according to the sign of the corresponding element of the hat matrix, and are weighted by the absolute value of the hat matrix element. (c) The random walk on the aggregate network in (a) for a walker starting at node 1 and finishing at node 2; edges are labelled by the associated transition probabilities. }
    \label{fig:agg-rw-flow}
\end{figure}

The aggregate model can be represented by an `aggregate network' where $w_{ab}$ is the weight associated with the edge $ab$.  We collect the aggregate edge weights in a $K\times K$ diagonal matrix, $\boldsymbol{W}=\mathrm{diag}(w_{ab})$.  Figure \ref{fig:agg-rw-flow} (a) shows a fictional example of an aggregate network with five treatments $a=1,2,3,4,5$. The aggregate weight matrix for this example is $\boldsymbol{W}=\mathrm{diag}(1,3,4,6,5,2,7)$. We write $\boldsymbol{B}$ for the $K\times N$ edge-incidence matrix of the aggregate network. Each column of $\boldsymbol{B}$ corresponds to a treatment in the network and each row corresponds to an edge.  To construct the matrix, one of the two treatments in each edge is designated as the `baseline treatment' for this edge without loss of generality. Entries are $+1$ in the column corresponding to the `baseline' treatment of the comparison represented by that row, and $-1$ in the column corresponding to the treatment compared to that baseline. For the example in Figure \ref{fig:agg-rw-flow} (a) the edge incidence matrix can be chosen as
\begin{align}
    \boldsymbol{B} = \begin{pmatrix}
    1 & -1 & 0 & 0 & 0 \\
    1 & 0 & -1 & 0 & 0 \\
    1 & 0 & 0 & 0 & -1 \\
    0 & 1 & -1 & 0 & 0 \\
    0 & 1 & 0 & -1 & 0 \\
    0 & 0 & 1 & 0 & -1 \\
    0 & 0 & 0 & 1 & -1
    \end{pmatrix},
\end{align}
where the columns represent treatments 1, 2, 3, 4, and 5, and the rows represent the edges (direct comparisons) 1-2, 1-3, 1-5, 2-3, 2-4, 3-5, and 4-5. In the following we will use this hyphenated notation when we refer to specific comparisons (e.g. 1-2 for the comparison between treatments 1 and 2). When we refer to a comparison between unspecified treatments $a$ and $b$, then we will use the notation $ab$, to avoid confusion with `$a$ minus $b$'. 

\subsection{Hat matrix and network estimates}
The \textit{network} estimates of the relative treatment effects $\hat\theta_{ab}^{\mathrm{net}}$ are obtained via
\begin{align}
\label{eq:NET}
    \boldsymbol{\hat\theta}^{\mathrm{net}} = \boldsymbol{H} \boldsymbol{\hat\theta}^{\mathrm{dir}},
\end{align}
where the hat matrix associated with the aggregate model is \cite{Rucker:2014}
\begin{align}
\label{eq:HAT}
    \boldsymbol{H} = \boldsymbol{B}(\boldsymbol{B}^\top \boldsymbol{W} \boldsymbol{B})^+\boldsymbol{B}^\top \boldsymbol{W}.
\end{align}
The hat matrix has dimension $K\times K$ where each row and each column correspond to one edge. We denote the element in the $ab$ row and $cd$ column by $H_{cd}^{(ab)}$. The matrix $\boldsymbol{L}=\boldsymbol{B}^\top \boldsymbol{W} \boldsymbol{B}$, with dimensions  $N\times N$ and rank $N-1$, is the Laplacian of the aggregate network. The matrix $\boldsymbol{L}^{+}=(\boldsymbol{B}^\top \boldsymbol{W} \boldsymbol{B})^{+}$ is its pseudo-inverse\cite{Rucker:2012, Gutman:2004}. The hat matrix describes how the direct evidence combines to give the network estimates. Each network estimate is a weighted linear combination of direct and indirect evidence. The coefficients of the estimates $\boldsymbol{\hat{\theta}}^{\mathrm{dir}}$ for each network treatment effect are found in the corresponding row of $\boldsymbol{H}$. The diagonal elements of $\boldsymbol{H}$ give the coefficients for the direct evidence while the off-diagonal elements indicate the contribution of indirect evidence. The larger the diagonal elements, the more weight is given to direct evidence \cite{konig:2013}. 

\subsection{Evidence flow}
\label{flow}
K{\"o}nig et al (2013) \cite{konig:2013} noted that each row in the hat matrix can be interpreted as a flow network. Focusing on one row of the hat matrix, the magnitude of the flow of evidence between two nodes is given by the absolute value of the element in the corresponding column of $\boldsymbol{H}$. The direction is determined by the sign of the element of the hat matrix. For the $ab$-row of the hat matrix one defines evidence flows $f^{(ab)}_{cd}$ (from $c$ to $d$) and $f^{(ab)}_{dc}$ (from $d$ to $c$) as follows\cite{konig:2013}:
 \begin{eqnarray}\label{eq:f}
 &&\mbox{if $H^{(ab)}_{cd}>0:$} ~~~ f_{cd}^{(ab)}=H_{cd}^{(ab)},~~~ f_{dc}^{(ab)}=0, \nonumber \\
&&  \mbox{if $H^{(ab)}_{cd}<0:$} ~~~ f_{cd}^{(ab)}=0,~~~~~~~~~ f_{dc}^{(ab)}=|H_{cd}^{(ab)}|. 
 \end{eqnarray}
Flows are non-negative, and only one of  $f^{(ab)}_{cd}$ and $f^{(ab)}_{dc}$ is non-zero.
 
It is important to note that each comparison $ab$ gives rise to a separate network of flows. We refer to these graphs as `evidence flow networks'. Due to the properties of the hat matrix each of these evidence flow networks is directed and acyclic. Specifically, in the network corresponding to the comparison $ab$, node $a$ only has outgoing edges, and node $b$ only incoming edges. The flow network then has the following properties:
\begin{enumerate}
    \item[1.] The total outflow from $a$ is equal to one, $\sum_{x}f_{ax}^{(ab)}=1$;
    \item[2.] the sum of inflows to node $b$ is also one, $\sum_{x}f_{xb}^{(ab)}=1$;
    \item[3.] and at every intermediate node, $c\neq a,b$, the sum of outflows equals the sum of inflows, $\sum_{x}f_{cx}^{(ab)}=\sum_{x}f_{xc}^{(ab)}$.
\end{enumerate}
 These properties were stated in Reference\cite{konig:2013}, and an algebraic proof for the first and the second property was given in Reference \cite{Papakon:2018}. We provide a heuristic argument for all three properties in Appendix~\ref{sec:heuristic}. 

Figure \ref{fig:agg-rw-flow} (b) shows the evidence flow network for the comparison 1-2 for the aggregate network in Figure \ref{fig:agg-rw-flow} (a).

\section{NMA, electrical networks and random walks}
\label{analogies}

In this section we set up the analogies between NMA, electrical networks and random walks. A summary of these analogies can be found in Table \ref{Tab:analogies}. 

\subsection{NMA and electrical networks}
\label{NMA_electric}

\begin{table}
\caption{~  Summary of the analogy between NMA, electrical networks and random walks (RW) on the aggregate network. } 
\centering
\begin{tabularx}{\linewidth}{| X | X | X |}
\hline
NMA & Electric circuit & RW on the aggregate network \\
\hline
&&\\
Treatments $1, 2, \hdots, N$ & Nodes $1,2,\hdots, N$ &  Nodes $1,2,\hdots, N$ \vspace{5pt} \\
Direct treatment comparisons & Edges (conducting wires) & Edges (along which a random walker can travel in both directions) \vspace{5pt}\\
${w}_{ab}$ inverse-variance weight associated with edge $ab$ on the aggregate network & $C_{ab} = R_{ab}^{-1}$ conductance (inverse resistance) across edge $ab$ &  $T_{ab} =C_{ab}/\sum_{c \neq a} C_{ac}= {w}_{ab}/\sum_{c\neq a}  {w}_{ac}$ probability that  a walker at node $a$ hops to node $b$ in the next step \vspace{5pt}\\
The aggregate hat matrix element $H^{(ab)}_{cd}$ that defines the flow of evidence through the direct comparison $cd$ for the network treatment effect $ab$  & Flow of current through edge $cd$ when a battery is attached across nodes $a$ and $b$ such that a unit current flows into $a$ and out of $b$ & Expected net number of times a walker starting at $a$ and ending at $b$ crosses the edge from $c$ to $d$\\
&&\\
\hline
\end{tabularx}
\label{Tab:analogies}
\end{table}

The connection between meta-analytic and electrical networks was first introduced by one of us \cite{Rucker:2012}. In the meta-analytic network, treatments are nodes connected by edges representing pairwise comparisons. On the other hand, edges in an electrical network represent resistors that connect at the nodes. In electric networks one then assigns an electric potential to each node, resulting in voltages (=differences in electric potential) across all edges. This in turn induces currents across the edges (current=voltage divided by resistance). Currents may also flow into/out of each node from/to the exterior to guarantee Kirchhoff's current law\cite{Urbano:2019} at each node. These external currents occur, for example, when a voltage source (battery) is attached to a pair of nodes.

The analogy between NMA and electric networks is based on the observation that resistances in parallel and sequential electrical circuits combine in the same way as variances of treatment effects in an NMA. Variance therefore corresponds to resistance. One can show that relative treatment effects are the analogue of voltages measured across edges, and weighted treatment effects the analogue of electrical current (see R\"ucker (2012)\cite{Rucker:2012} for details). This allows one to use graph theoretical tools, routinely applied to electrical networks, to address questions in NMA.

In R\"ucker (2012)\cite{Rucker:2012}, no voltages or external currents are applied directly to the electric circuit representing the NMA network. Instead, the starting point is given by (potentially) inconsistent measurements of treatment effects (voltages) across the edges of the network. It is then shown that finding the NMA estimates of treatment effects corresponds to finding the set of consistent voltages across all edges minimising the (Euclidian) distance to the inconsistent measurements.

Here, we extend this analogy and show that the elements of the hat matrix have an interpretation in the electric-circuit picture. More precisely, the elements of the row in the hat matrix corresponding to the comparison between treatments $a$ and $b$ can be obtained as follows: Connect a battery to nodes $a$ and $b$ in the electric circuit so that one unit of current flows from the exterior into node $a$, and out of the network (to the exterior) from node $b$. The external currents into/out of all other nodes are maintained at zero. This induces currents across the edges in the network. Our main result is then the following: The current along edge $cd$ is identical to the hat matrix element $H^{(ab)}_{cd}$. A detailed mathematical proof can be found in Appendix \ref{App:Current_Flow}.

\begin{figure}
    \centering
    \includegraphics[width=0.95\linewidth]{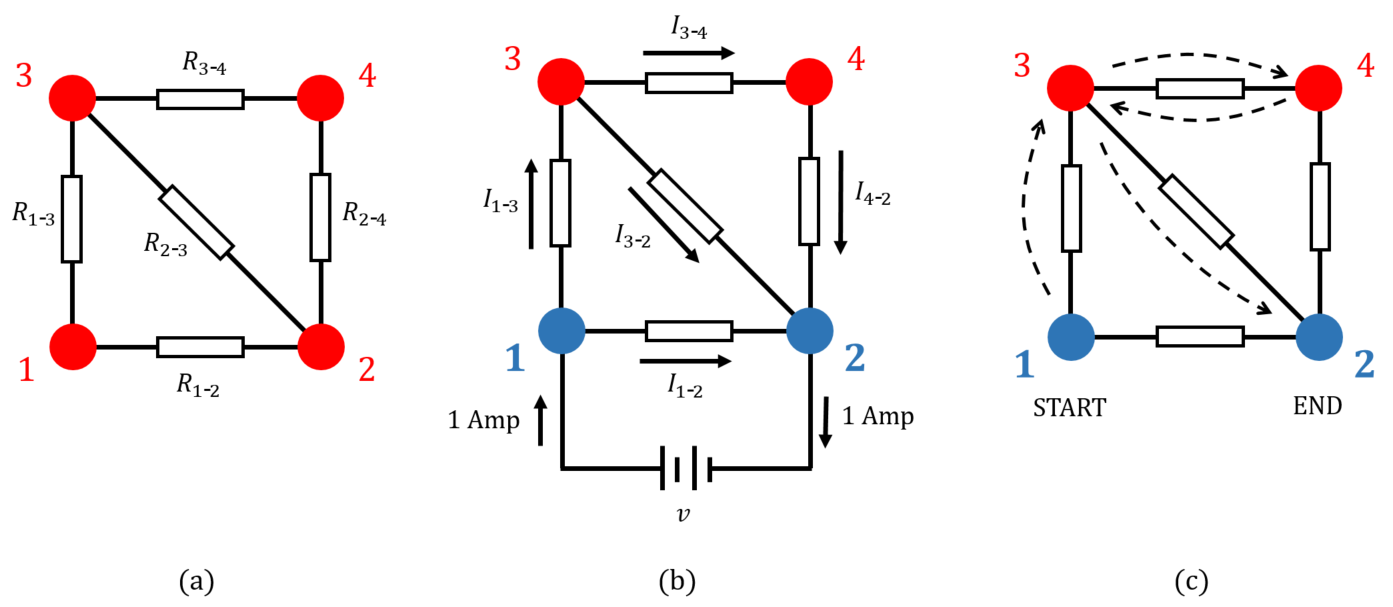}
    \caption{~An illustration of the interpretation of current. (a) An electrical network with associated edge resistances. (b) The same network with a battery attached across the edge 1-2 such that a unit current flows into 1 and out of 2. The current in edge $cd$ is labelled $I_{cd}$. Current is measured in amp{\'e}res, hence the unit current is labelled as `1 Amp'. The direction of the current induced in the edges is shown. (c) A possible path  taken by a random walker starting at node 1 and stopping at node 2. The sequence of nodes visited is $1\xrightarrow{}3\xrightarrow{}4\xrightarrow{}3\xrightarrow{}2$. For this particular realisation of the random walk, the net number of times the walker crosses edges 1-3 and 3-2 is one, while all other edges are crossed net zero times. The expected net number of times the walker crosses an edge is given by the currents shown in (b) for that edge \cite{Doyle:2000}. The focus on the comparison of nodes 1 and 2 in panels (b) and (c) is indicated by the blue colour of these nodes. }
    \label{fig:RW-electric}
\end{figure}

We illustrate this with a simple network of four nodes in Figure~\ref{fig:RW-electric}. Panel (a) shows a generic electrical circuit resulting from a meta-analytic graph with four treatment options and with direct comparisons between all pairs of treatments except treatments 1 and 4. We focus on the row in the hat matrix corresponding to the comparison between treatments $1$ and $2$. Using Equation~(\ref{eq:NET}) we have for this example
\begin{align}
\label{eq:linear_hat2}
    \hat{\theta}^{\mathrm{net}}_{\text{1-2}} = H_{\text{1-2}}^{(\text{1-2})}\hat{\theta}^{\mathrm{dir}}_{\text{1-2}} + H_{\text{1-3}}^{(\text{1-2})}\hat{\theta}^{\mathrm{dir}}_{\text{1-3}}  + H_{\text{2-3}}^{(\text{1-2})}\hat{\theta}^{\mathrm{dir}}_{\text{2-3}}  + H_{\text{2-4}}^{(\text{1-2})}\hat{\theta}^{\mathrm{dir}}_{\text{2-4}} + H_{\text{3-4}}^{(\text{1-2})}\hat{\theta}^{\mathrm{dir}}_{\text{3-4}}.    
\end{align}
Our result indicates that the coefficients $H^{(\text{1-2})}_{cd}$ can be obtained from the setup shown in Figure~\ref{fig:RW-electric}. A battery is attached to nodes $1$ and $2$ and the voltage of the battery is chosen such that one unit of current flows into node $1$ (from the battery) and out of node $2$ (into the battery). This induces currents in the five edges (resistors) of the electric circuit. These currents are the hat matrix elements in Equation~(\ref{eq:linear_hat2}). Via Equation~(\ref{eq:f}) these then determine the flow of evidence.

\subsection{Electrical networks and random walks}
\label{electric_RW}
\subsubsection{Definitions and notation}

\begin{figure}
    \centering
    \includegraphics[width=0.5\linewidth]{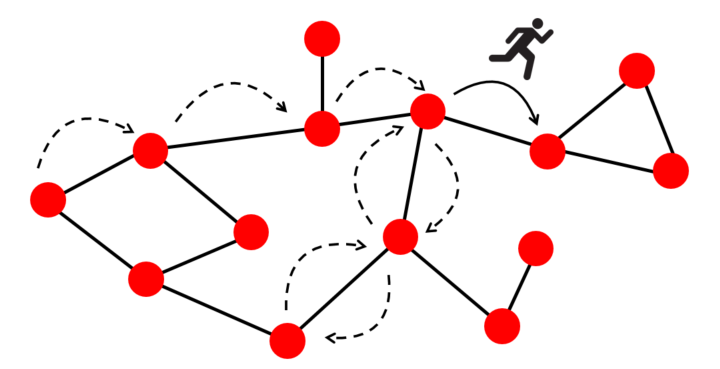}
    \caption{~An illustration of a random walker moving on a network graph. The walker starts its journey from the far left node. The arrows show the path taken by the walker for one realisation of the random walk. The figure indicates the `current' position of the walker as it hops between two nodes. The solid arrow indicates this transition. The dotted arrows indicate the previous transitions made between nodes by the walker.  }
    \label{fig:RW}
\end{figure}

As illustrated in Figure \ref{fig:RW}, a random walk on a graph is a stochastic process consisting of succession of ‘hops’ between neighbouring nodes (nodes connected by an edge). We use the word `path' to describe the sequence of nodes visited by the walker, including repeat visits to individual nodes.  We always assume that time is discrete. The walk is then a Markov process described by an $N\times N$ transition matrix, $\boldsymbol{T}$, where $N$ is the number of nodes in the network. The element $T_{ab}$ of this matrix is the probability that a walker, currently at node $a$, moves to node $b$ in the next time step. These probabilities only depend on the current position of the walker, and not on the path taken to reach that position. One has $\sum_b T_{ab}=1$ for all $a$, i.e.  $\boldsymbol{T}$ is a stochastic matrix. 

The connection between random walks and electrical networks has been recognised for some time \cite{Kakutani:1945, Kemeny:1966, Kelly:1979} and is described extensively in Doyle and Snell (2000) \cite{Doyle:2000}. Here we will only summarise the concepts and known results that are most relevant for our work. 

Starting from an electrical network with given resistances $R_{ab}$ a random walk process can be constructed by defining the transition probabilities ($a\neq b$)
\begin{align}
\label{eq:Pxy}
    T_{ab} =  \frac{R_{ab}^{-1}}{\sum_{c\neq a}R_{ac}^{-1}}.
\end{align}
This definition indicates that transitions from one node to another occur in proportion to the inverse resistance of the direct connection between the two nodes (if there is no direct connection, then no hop can occur between the two nodes). We set $T_{aa}=0$ for all $a$. The denominator in Equation~(\ref{eq:Pxy}) ensures normalisation ($\sum_b T_{ab}=1$). 

We always assume the network does not divide into multiple disconnected components. As a result, the transition matrix defined in Equation (\ref{eq:Pxy}) is such that a walker starting at any node $a$ will eventually reach any other node $b\neq a$ with finite probability.

\subsubsection{Interpretation of electrical current}\label{sec:interpret_current}
Electrical current can be interpreted in the random-walk picture  as follows\cite{Doyle:2000}: 
When a voltage is applied between two nodes $a$ and $b$ such that the total current flowing into $a$ and out of $b$ from the exterior is 1, the current induced in each edge, $cd$, is equal to the expected \textit{net} number of times a random walker, starting at $a$ and walking until it reaches $b$, moves along the edge from $c$ to $d$. The net number of times the walker moves from $c$ to $d$ is the number of crossings in the direction from $c$ to $d$ minus the number of crossings in the opposite direction.

To describe this mathematically we need to ensure that no more hops occur when the walker reaches the designated end point $b$. In other words, this node must become absorbing. This is achieved by setting the elements $T_{bc}$ to zero for all $c$. For later convenience we denote the resulting modified transition matrix by $\boldsymbol{T}^{(ab)}$, recognising that the modifications made only depend on the choice of $b$, and not on $a$. Mathematically, we have $T^{(ab)}_{bc}=0$ for all $c\neq b$, and  $T^{(ab)}_{cd}=T_{cd}$ for $c\neq b$ and all $d$. We set $T^{(ab)}_{bb}$ to unity.

Now consider random walks starting at node $a$ and then following the process defined by the transition matrix $\boldsymbol{T}^{(ab)}$. All walks therefore end at node $b$. The probability that a walker takes a particular path $\pi$ connecting $a$ and $b$ can be written as 
\begin{align}
\label{eq:P_pi}
    P^{(ab)}(\pi) = \prod_{\{xy \in \pi\}} T^{(ab)}_{xy},
\end{align}
where the notation $\{xy \in \pi\}$ indicates the set of pairs of successive nodes in the path $\pi$. We note that $P^{(ab)}(\pi)$ is non-zero if and only if the path $\pi$ starts at $a$ and ends when $b$ is reached for the first time.

The average number of net crossings from node $c$ to node $d$ along paths starting at $a$ and ending at $b$ can therefore be obtained as
\be\label{eq:nbar}
\overline{N_{cd}^{(ab)}}=\sum_{\pi} P^{(ab)}(\pi) N_{cd}(\pi),
\ee
where $N_{cd}(\pi)$ is the net number of crossings from $c$ to $d$ along path $\pi$. We note that this quantity can be negative; this occurs if the walker makes more transitions from $d$ to $c$ than from $c$ to $d$. The sum in Equation~(\ref{eq:nbar}) extends over all paths connecting $a$ and $b$.

To develop some intuition, consider again the electrical network in Figure \ref{fig:RW-electric} (a). Assume that we are interested in the scenario where the external current flows into node 1 and out of node 2, but not into or out of any of the other nodes. We then start the random-walk process at node 1, and use transition probabilities as defined in Equation~(\ref{eq:Pxy}) until the walker reaches node $2$. In the first step, the walker either hops to node 2 (this occurs with probability $T_{1\text{-}2}$) or to node 3 (with probability $T_{1\text{-}3}$). If the walker hops to node 2, the walk stops and the path taken by the walker is 1 $\xrightarrow{}$2. Otherwise, the walker is at node 3 and in the next step it can transition to 2, 4 or back to 1 with respective probabilities $T_{\text{3-2}}$, $T_{\text{3-4}}$ and $T_{\text{3-1}}$. This process continues until the walker eventually reaches node 2. The current through the edge $cd$ is then given by the expected net number of times such a walker crosses the edge from $c$ to $d$ before it arrives at node 2. A crossing in the direction from $d$ to $c$ contributes negatively to this value.

Since the random walker can move in both directions along the network edges, there are infinitely many paths the walker can take as it travels from node 1 to node 2 in this example. Figure \ref{fig:RW-electric} (c) shows one possible path, 1 $\xrightarrow{}$3 $\xrightarrow{}$4 $\xrightarrow{}$3 $\xrightarrow{}$2.  The probability the random walker takes this path is given by the product of the individual transition probabilities along the path, that is
\begin{align}
    P^{\text{(1-2)}}(1\xrightarrow{}3 \xrightarrow{} 4 \xrightarrow{} 3 \xrightarrow{} 2) =T_{\text{1-3}}T_{\text{3-4}}T_{\text{4-3}}T_{\text{3-2}}. 
\end{align}
Although $P^{(ab)}(\pi)$ can be obtained relatively easily for each path $\pi$, carrying out the sum in Equation~(\ref{eq:nbar}) by exhaustive enumeration of all relevant paths is not practicable. This is because there are generally infinitely many paths starting and ending at the designated nodes (due to the possibility to hop back to nodes visited earlier).

The analogy between electrical circuits and random walks\cite{Doyle:2000} however can be be used to calculate the expected number of net crossings through an edge analytically. This is detailed in Appendices  \ref{App:RW_electric} and \ref{App:CalcRWFlow}, see in particular Equation~(\ref{eq:fN}).

The expected number of net crossings can also be obtained from simulations of the random-walk process. An ensemble of walkers is released at the starting point $a$. Each walker then independently hops from node to node on the network with transition rates as in Equation~(\ref{eq:Pxy}) until it hits the designated endpoint (node $b$). The process then stops. For each walker the net number of crossings from $c$ to $d$ can be recorded, and this is then averaged over the ensemble of walkers.

\subsection{Random walk on a meta-analytic network}
\label{RW_NMA}

As described above, conductance (inverse resistance) in an electrical network has an analogue in terms of both NMA, and random walks. Exploiting these analogies, we now define a random-walk process on a meta-analytic network via the transition rates 
\begin{align}
\label{eq:Pxy-V}
    T_{ab} = \frac{{w}_{ab}}{\sum_{c\neq a} {w}_{ac}},
\end{align}
with weights $w_{ab}$ associated with the edges as discussed in Section~\ref{sec:aggregate}, see in particular Equation~(\ref{eq:weight}).

In order to study walks starting at node $a$ and ending at $b$ we use the matrix $\boldsymbol{T}^{(ab)}$ as defined in Section~\ref{sec:interpret_current}. This enforces absorption of the walker at node $b$ when this node is reached. For the example aggregate network in Figure \ref{fig:agg-rw-flow} (a), the transition matrix for a random walk starting at node 1 and ending at node 2 is
\begin{align}
\label{eq:T12_eg}
    \boldsymbol{T}^{(\text{1-2})} = \begin{pmatrix}
    0 & 1/8 & 3/8 & 0 & 4/8 \\
    0 & 1 & 0 & 0 & 0 \\
    3/11 & 6/11 & 0 & 0 & 2/11 \\
    0 & 5/12 & 0 & 0 & 7/12 \\
    4/13 & 0 & 2/13 & 7/13 & 0 \\
    \end{pmatrix}.
\end{align}
Each row and column of $\boldsymbol{T}^{(\text{1-2})}$ represents a treatment in the network, $a=1, 2, 3, 4, 5$. Given that we focus on the comparison between treatments 1 and 2, node 1 is the start point of the walk, and node 2 is absorbing. Therefore, the row corresponding to treatment 2 contains only zeroes except for the diagonal element which is equal to one (when the walker reaches node $2$ it stays there indefinitely). The entries in each row of the matrix in Equation (\ref{eq:T12_eg}) sum to one. The diagonal elements of $\boldsymbol{T}^{(\text{1-2})}$ (except for the element relating to node 2) are zero. This indicates that, with the exception of the absorbing state, the random walker cannot stay at the same place at any step. Figure \ref{fig:agg-rw-flow} (c) illustrates the dynamics of the random walk from node 1 to node 2 for this example. 

In Section \ref{NMA_electric} we made the connection between the flow of electric current and the flow of evidence in an NMA. Using the interpretation of current as a random walk we can now establish the following analogy: For the comparison of treatments $a$ and $b$, the hat matrix element $H_{cd}^{(ab)}$ that defines the flow of evidence  through the direct comparison $cd$ is equal to the expected \textit{net} number of times a random walker starting at node $a$ on the aggregate NMA network moves along the edge from $c$ to $d$ before it reaches node $b$. In other words, we equate 
\begin{align} H_{cd}^{(ab)} = \overline{N_{cd}^{(ab)}},\end{align} 
and define the flow of evidence $f_{cd}^{(ab)}$ in terms of $H^{(ab)}_{cd}$ via Equation (\ref{eq:f}). 

In summary, we have used existing analogies between electric circuits and random walks on the one hand, and network-meta analysis and electric circuits on the other to introduce an interpretation of the flow of evidence in network meta-analysis in terms of random walks. The analogies between all three areas are highlighted in Table \ref{Tab:analogies}.

\section{Proportion contribution}
\label{prop}

In this section we present a random-walk interpretation and construction of the so-called `proportion contribution matrix' \cite{Papakon:2018}. While the general idea is similar to the random-walk approach to evidence flow in the previous section, it is important to note that the random walk now no longer takes place on the meta-analytic network. Instead, walkers move on the evidence flow network. As explained in more detail below, the entries of the proportion contribution matrix in NMA can then be obtained from this random walk.

We show that the random-walk approach overcomes the limitations of the algorithm proposed for the evaluation of proportion contributions in Papakonstantinou et al (2018) \cite{Papakon:2018}. In particular, it provides an analytical expression for proportion contributions, removing ambiguity and reducing the computational effort required. Furthermore, unlike the numerical algorithm, the random-walk approach identifies all paths of evidence so that all potential sources of bias are taken into account. In Section \ref{prop_1} we introduce the concept of proportion contributions. In \ref{Algorithm} we describe the algorithm in Papakonstantinou et al and its limitations. We then present and discuss the random-walk approach in Section \ref{prop_RW}. 

\subsection{Background and  definition}
\label{prop_1}
In NMA it is important to assess the influence of individual study bias on the estimates obtained from the network. To this end, the CINeMA framework and software \cite{Nikola:2020, Papakon:2020} provides a user friendly system to assess confidence in the results from an NMA. One function of the software is to display the relative influence of evidence that comes from studies with high, moderate and low risk of bias on each network treatment effect. This assessment involves calculating the matrix of so-called `proportion contributions' \cite{Papakon:2018}. This matrix describes how much each direct treatment effect contributes to each network treatment effect as a relative \textit{proportion}. The idea of the proportion contribution matrix is based on the hat matrix. The elements of the hat matrix are the coefficients of the linear relation between network estimates and direct estimates in the NMA as described in Equation~(\ref{eq:NET}). These coefficients can be positive or negative. The proportion contribution matrix uses the properties of the hat matrix and translates the elements of $\boldsymbol{H}$ to positive proportion contributions, where the total contribution is normalised to one. We now explain this in more detail using the work of Papakonstantinou et al (2018) \cite{Papakon:2018}. 

\begin{figure}
    \centering
    \includegraphics[width=0.95\linewidth]{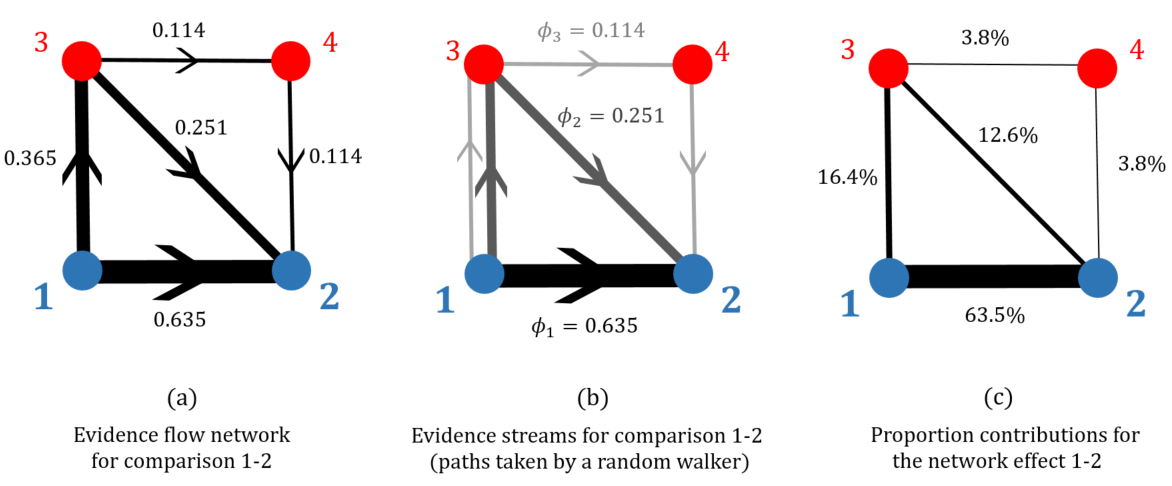}
    \caption{~Illustration of evidence flow, streams of evidence and proportion contributions for a network of topical antibiotics without steroids for chronically discharging ears presented in Macfadyen (2005) \cite{Macfadyen:2005}. Node 1 is no treatment; 2 is quinolone antibiotic; 3 is antiseptic; and 4 is non-quinolone antibiotic. (a) The evidence flow network for comparison 1-2, based on Figure 1, panel (b) in Papakonstantinou et al (2018)\cite{Papakon:2018}. The edge labels are the entries of the 1-2 row of the hat matrix, their signs are associated with the direction of the arrows. (b) The decomposition of edge flows into flow through paths of evidence as estimated by the algorithm in Papakonstantinou et al. The paths of evidence shown are equivalent to the possible paths taken by a random walker on the evidence flow network. (c) The proportion contributions (expressed as percentages) of each direct treatment effect to the network estimate of the 1-2 relative treatment effect. }
    \label{fig:PROP}
\end{figure}

Consider the example network in Figure \ref{fig:PROP} (a). This relates to an NMA of the four topical antibiotics given in the figure caption for the treatment of chronically discharging ears \cite{Macfadyen:2005}. To keep the text concise we label the treatments 1, 2, 3 and 4. In accordance with Equation (\ref{eq:NET}), the network estimate of comparison 1-2 is given by the linear equation (\ref{eq:linear_hat2}), which we repeat here for clarity,
\begin{align}
\label{eq:linear_hat}
    \hat{\theta}^{\mathrm{net}}_{\text{1-2}} = H_{\text{1-2}}^{(\text{1-2})}\hat{\theta}^{\mathrm{dir}}_{\text{1-2}} + H_{\text{1-3}}^{(\text{1-2})}\hat{\theta}^{\mathrm{dir}}_{\text{1-3}}  + H_{\text{2-3}}^{(\text{1-2})}\hat{\theta}^{\mathrm{dir}}_{\text{2-3}}  + H_{\text{2-4}}^{(\text{1-2})}\hat{\theta}^{\mathrm{dir}}_{\text{2-4}} + H_{\text{3-4}}^{(\text{1-2})}\hat{\theta}^{\mathrm{dir}}_{\text{3-4}}.    
\end{align}
We can think of the expression on the right-hand side as a combination of different direct and indirect estimates of $\theta_{\text{1-2}}$. The direct estimate is simply $\hat{\theta}^{\mathrm{dir}}_{\text{1-2}}$. We obtain one indirect estimate using node 3 and the consistency equation,
\begin{align}
    \hat{\theta}_{\text{1-2}}^{\mathrm{ind}(1)} = \hat{\theta}^{\mathrm{dir}}_{\text{1-3}} - \hat{\theta}^{\mathrm{dir}}_{\text{2-3}}.
\end{align}
A second indirect estimate is found via nodes 3 and 4,
\begin{align}
    \hat{\theta}_{\text{1-2}}^{\mathrm{ind}(2)} = \hat{\theta}^{\mathrm{dir}}_{\text{1-3}} - (\hat{\theta}^{\mathrm{dir}}_{\text{2-4}} - \hat{\theta}^{\mathrm{dir}}_{\text{3-4}}).
\end{align}
These three ways of estimating $\theta_{\text{1-2}}$ correspond to so-called `paths of evidence' on the evidence flow network\cite{Papakon:2018}. We label these paths $\pi_i$ ($i=1,2,3$). As illustrated in Figure \ref{fig:PROP} (b), these are $\pi_1=1\xrightarrow{}2$, $\pi_2=1\xrightarrow{}3\xrightarrow{}2$, and $\pi_3=1\xrightarrow{}3\xrightarrow{}4\xrightarrow{}2$. We can now write the network estimate $\hat{\theta}^{\mathrm{net}}_{\text{1-2}}$ as a linear combination of the estimates $\hat{\theta}^{\mathrm{dir}}_{\text{1-2}}$, $\hat{\theta}_{\text{1-2}}^{\mathrm{ind}(1)}$, and $\hat{\theta}_{\text{1-2}}^{\mathrm{ind}(2)}$. That is,
\begin{align}
    \hat{\theta}^{\mathrm{net}}_{\text{1-2}} &= \phi_1 \hat{\theta}^{\mathrm{dir}}_{\text{1-2}} + \phi_2 \hat{\theta}_{\text{1-2}}^{\mathrm{ind}(1)} + \phi_3 \hat{\theta}_{\text{1-2}}^{\mathrm{ind}(2)}\nonumber \\
    &= \phi_1 \hat{\theta}^{\mathrm{dir}}_{\text{1-2}} + \phi_2(\hat{\theta}^{\mathrm{dir}}_{\text{1-3}} - \hat{\theta}^{\mathrm{dir}}_{\text{2-3}}) + \phi_3(\hat{\theta}^{\mathrm{dir}}_{\text{1-3}} - \hat{\theta}^{\mathrm{dir}}_{\text{2-4}} + \hat{\theta}^{\mathrm{dir}}_{\text{3-4}}). \label{eq:linear_phi}
\end{align}
The coefficients, $\phi_i$, define the flow of evidence through each path $\pi_i$, see Papakonstantinou et al (2018) \cite{Papakon:2018}.

Figure \ref{fig:PROP} (b) shows how the flows in each edge, described by the hat matrix coefficients, are deconstructed into the flows through each path of evidence, described by the coefficients $\phi_i$. In this example, only the edge 1-3 is used for more than one path. When calculating the flow through each path, the flow in edge 1-3 is `split' between the paths $\pi_2=1\xrightarrow{}3\xrightarrow{}2$, and $\pi_3=1\xrightarrow{}3\xrightarrow{}4\xrightarrow{}2$ according to the flow in the subsequent edges along those two paths.

A so-called `stream' of evidence\cite{Papakon:2018} is a pair consisting of a path and the flow associated with this path, $S_i=(\pi_i, \phi_i)$. The proportion contribution of each direct comparison $cd$ to the network estimate of each comparison $ab$, is then defined as \cite{Papakon:2018} 
\begin{align}
\label{eq:PropCont}
    p^{(ab)}_{cd} = \sum_{i:~ cd \in \pi_i} \frac{\phi_i}{|\pi_i|},
\end{align}
where $|\pi_i|$ is the number of edges that make up the path. The sum extends over all paths in the evidence flow network for the comparison $ab$ that contain the edge $cd$. We note that all such paths start at $a$ and end at $b$, and, because the evidence flow network is acyclic, multiple visits to the same node do not occur.

For simple examples, such as the one in Figure \ref{fig:PROP}, one can obtain the path flows $\phi_i$ by directly comparing coefficients in Equations (\ref{eq:linear_hat}) and  (\ref{eq:linear_phi}). Using the properties of the hat matrix in Section~\ref{flow} one can then also see that $\phi_i\geq 0$ for all $i$, and that $\sum_i \phi_i=1$. This means that the proportion contributions in Equation (\ref{eq:PropCont}) are also non-negative, and sum to one. Figure~\ref{fig:PROP}(c) shows the proportion contributions, expressed as percentages, for the example in Figure \ref{fig:PROP} (a).

For larger, more connected networks it is not immediately clear how to obtain the $\phi_i$. In particular, when there are more paths than edges, expressing the $\phi_i$ in terms of the coefficients of the hat matrix is non-trivial. Papakonstantinou et al\cite{Papakon:2018} present an iterative algorithm to identify streams for a general evidence flow network. We will now briefly describe this.

\subsection{Existing iterative numerical algorithm to determine streams of evidence}
\label{Algorithm}
 Broadly speaking, each iteration of the algorithm consists of the following steps: (i) A path in the evidence flow network is selected. (ii) The  minimum flow through the edges making up the path is identified. This is assigned as the flow associated with the path. (iii) The flow of the path is subtracted from the values of flow in the edges that make up that path. This means that the edge corresponding to the minimum flow in that path is removed from the graph. (iv) A new path is then selected from the remaining graph. The process repeats until all the evidence flow in the edges has been assigned to a path.   

Different methods for selecting the paths in step (i) give rise to multiple variants of the algorithm. For example, paths may be selected at random or in order from shortest to longest. We refer to these approaches as `Random' and `Shortest' respectively. To deal with multiple paths of the same length, the Shortest algorithm assigns a cost to each path based on the evidence flow in each edge along the path \cite{Papakon:2018}. The algorithm then selects paths in order from smallest to largest cost.  Each time the Random algorithm is run it selects the paths in a different order and, potentially, gives a different outcome. For simple networks such as the example in Figure \ref{fig:PROP}, the order of selection does not affect the outcome. However, for more complicated networks this is not the case. In some graphs, the flow of evidence is fully assigned to streams before every possible path has been selected. The remaining paths can then not be associated with any flow. Critically, this approach means that many paths of evidence are not identified and their contribution (along with any potential bias) is not accounted for. The set of paths that are missed in this way can depend on the order in which paths are selected by the algorithm. Examples of this behaviour are presented in Supplementary File 3 in Papakonstantinou et al (2018)\cite{Papakon:2018} and in Section \ref{apply-prop} of this paper. 

One potential remedy consists of averaging results from the Random algorithm by  Papakonstantinou et al over a large number of realisations. We call this method `Average'. Provided enough realisations are generated the Average algorithm will eventually identify every evidence path. However, because of the nature of the algorithm, the number of times a particular path is sampled by this method can depend on features of the network not directly related to the path. In step (iii) of the algorithm the edge associated with the smallest flow in a particular path is removed from the network. This means that any other path containing this edge can no longer be selected. As a result, paths that do not share edges with any other paths will be selected in every run of the algorithm, whereas paths which do share edges with other paths will be sampled less often. It is therefore not clear how to interpret average proportion contributions determined in this way. Furthermore, this approach is computationally intensive as it relies on repeating the (already iterative) algorithm many times. For this reason, this version of the algorithm is not implemented in current software.

To overcome these limitations, we develop a random-walk approach for deriving the streams of evidence. We will now describe this.

\subsection{Random walk on the evidence flow network}
\label{prop_RW}

To obtain the evidence streams we define a random walk on the evidence flow network for comparison $ab$. We denote the transition matrix for this model by $\boldsymbol{U}^{(ab)}$ to distinguish it from the random-walk on the aggregate NMA network defined in Section~\ref{RW_NMA}. We note that there is a different evidence flow network for each treatment comparison $ab$. We indicate this by the superscript $(ab)$. Since the evidence flow network has directed edges the walker can only move in one direction along each edge (in the direction of evidence flow). Node $a$ in the evidence flow network for comparison $ab$ has only outgoing edges, and node $b$ only incoming edges. We also note that the evidence flow network is acyclic \cite{konig:2013}. This means that a walker can never visit any node more than once. 

It is important to distinguish carefully between the random-walk model on the aggregate network and that on the evidence flow network. In Section \ref{RW_NMA} we defined a transition matrix for a random walker moving from node $a$ to node $b$ on the aggregate meta-analytic network. The walker was allowed to move in both directions along the edges of the network. We labelled this transition matrix $\boldsymbol{T}^{(ab)}$ where the superscript indicates the start and end nodes of the walk, i.e., the treatment comparison we are interested in. By analysing the average movement of the walker, we obtained the evidence flow. In this section we focus instead on a random walk on the evidence flow network, and our aim is to construct streams of evidence. The two  approaches are summarised and contracted in Table \ref{Tab:RW}.

\begin{table}
\caption{~Summary of the two random-walk approaches to NMA. In one approach (`aggregate') the walker moves on the undirected aggregate network. In the second (`evidence flow'), the walker moves on the directed acyclic evidence flow network for a particular comparison of treatments. The transition matrices are denoted by $\boldsymbol{T}$ and $\boldsymbol{U}^{(ab)}$ respectively. Except for the imposition of a suitable absorbing state (see text) the transition probabilities on the aggregate network do not depend on the particular comparison that is studied. In contrast, there are separate evidence flow networks (and hence random-walk models) for each comparison $ab$, hence the superscript in $\boldsymbol{U}^{(ab)}$. The first column in the table indicates the sections in the text containing further definitions and details.}
\begin{tabularx}{\columnwidth}{| l | l | X | X | X |}
\hline
Section & Network  & Transition probabilities & Measured quantity & Outcome \\
\hline
~&~&~&~&~\\
\ref{RW_NMA} & Aggregate  &  $\begin{aligned}[t]T_{cd}= \frac{w_{cd}}{\sum\limits_{x\neq c} w_{cx}}\end{aligned}$ &  Expected net number of times a walker crosses an edge while travelling from $a$ to $b$&  Flow of evidence through the edge (elements of the hat matrix in the row corresponding to comparison $ab$)   \\
\ref{prop_RW} & Evidence flow   &  $\begin{aligned}[t] U_{cd}^{(ab)} &= \frac{H_{cd}^{(ab)}}{\sum\limits_{x\neq c} H_{cx}^{(ab)}}  \text{ if } H_{cd}^{(ab)}>0 \\
&=0  \hspace{28pt} \text{ if } H_{cd}^{(ab)}<0  \end{aligned}$ & Proportion of walkers taking a particular path while travelling from $a$ to $b$& Evidence streams for the comparison between $a$ and $b$ \\
~&~&~&~&~\\
\hline
\end{tabularx}
\label{Tab:RW}
\end{table}

To illustrate this, we consider the evidence flow network for comparison 1-2 in Figure \ref{fig:PROP} (a). We now construct a transition matrix for a random walk on this directed acyclic graph assuming that the walker starts at node 1. In contrast to random walks on the undirected meta-analytic graphs in Section \ref{RW_NMA}, the walker can only move in one direction across each edge as indicated by the direction of evidence flow. If the flow $f_{cd}^{(ab)}=0$ (because the associated hat matrix element $H_{cd}^{(ab)}\leq 0$), then no hop from $c$ to $d$ can occur. Each possible transition occurs with probabilities proportional to the evidence flows indicated in Figure~\ref{fig:PROP} (a). More generally, for the evidence flow network of comparison $ab$, the elements of the transition matrix $\boldsymbol{U}^{(ab)}$ are given by 
\begin{align}
\label{eq:U}
    U_{cd}^{(ab)} = \frac{f^{(ab)}_{cd}}{\sum_{x\neq c} f^{(ab)}_{cx}} =  \begin{cases}
    \frac{H_{cd}^{(ab)}}{\sum_{x\neq c} H_{cx}^{(ab)}} & \text{ if } H_{cd}^{(ab)}>0\\
    ~&~\\
    0 & \text{ if } H_{cd}^{(ab)}<0.
    \end{cases}
\end{align}
For the comparison $ab$, the walker remains at $b$ indefinitely once it gets there, i.e., we have $U_{bb}^{(ab)}=1$, and the probability of transitioning from $b$ to any other node $c\neq b$ is $U_{bc}^{(ab)}=0$. All other elements of the matrix $\boldsymbol{U}^{(ab)}$ are given by Equation~(\ref{eq:U}).  

For the example in Figure \ref{fig:PROP} (a), the transition matrix for a random walk on this graph is
\begin{equation}
\label{eq:egP-prop}
    \boldsymbol{U}^{(\text{1-2})} = \begin{pmatrix}
    0 & 0.635 & 0.365 & 0 \\
    0 & 1 & 0 & 0  \\
    0 & \frac{0.251}{0.251+0.114} & 0 & \frac{0.114}{0.251+0.114} \\
    0 & 1 & 0 & 0  
    \end{pmatrix} = \begin{pmatrix}
    0 & 0.635 & 0.365 & 0 \\
    0 & 1 & 0 & 0  \\
    0 & 0.688 & 0 & 0.312 \\
    0 & 1 & 0 & 0  
    \end{pmatrix}.
\end{equation}
The third row of $\boldsymbol{U}^{(\text{1-2})}$ corresponds to transitions from node 3. From Equation (\ref{eq:U}) and the edge flows shown in Figure \ref{fig:PROP} (a), we find that if the walker is at node 3, then it moves to either node 2 or node 4 with probabilities $0.251/(0.251+0.114)$ and $0.114/(0.251+0.114)$ respectively. Similar calculations are done to find the elements in the other rows. Once arrived at $2$ the walker remains there indefinitely. This behaviour is described by the second row of $\boldsymbol{U}^{(\text{1-2})}$. 

The walker can take one of three paths from 1 to 2: $\pi_1=1\xrightarrow{}2$, $\pi_2=1\xrightarrow{}3\xrightarrow{}2$, or $\pi_3=1\xrightarrow{}3 \xrightarrow{}4 \xrightarrow{} 2$. These are the same as the paths of evidence defined in Section \ref{prop_1} and are illustrated in Figure \ref{fig:PROP} (b). The probability of a walker taking a certain path is given by the product of the individual transition probabilities associated with each edge along that path (Equation (\ref{eq:P_pi})). For example, the probability that a random walker takes the path $1\xrightarrow{}3\xrightarrow{}2$ is $P^{(\text{1-2})}(\pi_2) = U^{(\text{1-2})}_{\text{1-3}}U^{(\text{1-2})}_{\text{3-2}}=0.365\times 0.688$. 

The probability that a walker takes a given path can also be measured from simulations of the random-walk process on the evidence flow network. To do this one simulates a large ensemble of independent walkers, and measures the proportions of walkers taking each path. We can think of this as flows of walkers through the different paths. We use this interpretation to provide a general analytical definition of the flow of evidence through a particular path: for the evidence flow network for comparison $ab$, we define
\begin{align}
\label{eq:phi_RW}
    \phi_i = P^{(ab)}(\pi_i) = \prod_{cd \in \pi_i} U_{cd}^{(ab)}.
\end{align}
With this definition we can construct decompositions such as the one in Equation (\ref{eq:linear_phi}) for all networks. From the $\phi_i$ the proportion contributions can then be calculated via Equation (\ref{eq:PropCont}).

For the example in Figure \ref{fig:PROP} (a), Equation (\ref{eq:phi_RW}) leads to the streams, 
\begin{align}
    &S_1=(\pi_1,\phi_1): & &\pi_1=1\xrightarrow{}2  &  &\phi_1 = U^{\text{(1-2)}}_{\text{1-2}} = 0.635\\
     &S_2=(\pi_2,\phi_2): & &\pi_2=1\xrightarrow{}3\xrightarrow{}2 &  &\phi_2 = U^{\text{(1-2)}}_{\text{1-3}}U^{\text{(1-2)}}_{\text{3-2}} = 0.365\times \frac{0.251}{0.251+0.114} = 0.251\\
     &S_3=(\pi_3,\phi_3): & &\pi_3=1\xrightarrow{}3 \xrightarrow{} 4 \xrightarrow{}2 & &\phi_3 = U^{\text{(1-2)}}_{\text{1-3}}U^{\text{(1-2)}}_{\text{3-4}}U^{\text{(1-2)}}_{\text{4-2}}=0.365\times \frac{0.114}{0.251+0.114} \times 1 = 0.114
\end{align}
For this simple example the random-walk approach results in the same evidence streams (and therefore proportion contributions) as the algorithm by Papakonstantinou et al, see Figure \ref{fig:PROP} (b).

The random-walk approach provides an analytical construction of the proportion contributions.  The outcome is  unambiguous and the method is computationally more efficient than the iterative numerical algorithm. In the following section we demonstrate how the random-walk approach can be used for the more intricate network from Section \ref{real_data}.

\pagebreak
\section{Application to real data set}
\label{apply}
\begin{figure}
    \centering
    \includegraphics[width=0.75\linewidth]{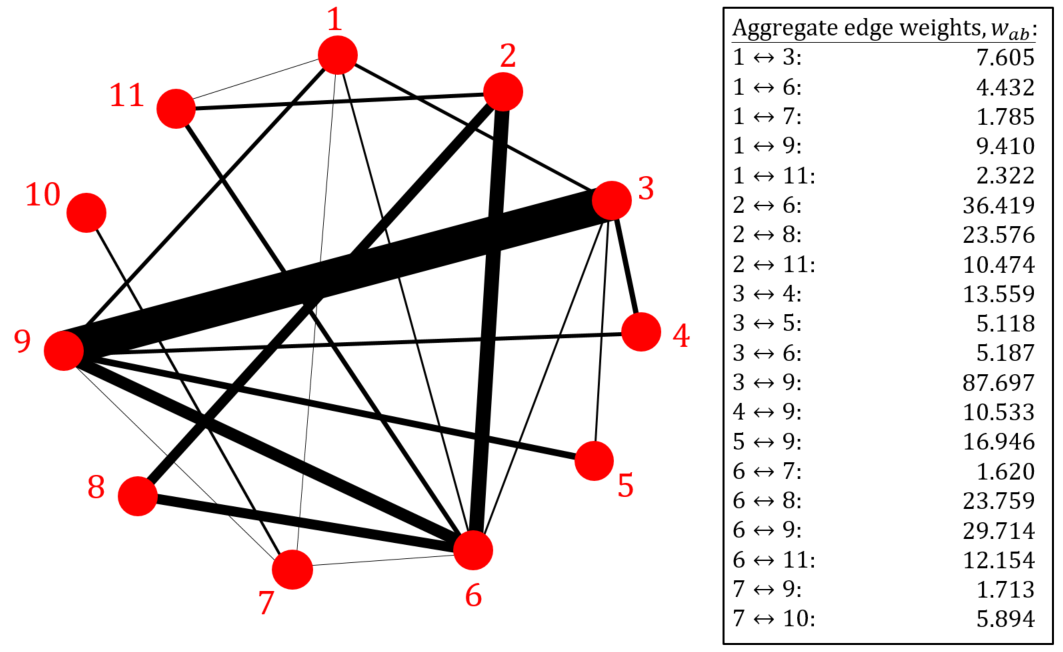}
    \caption{~The aggregate network for the depression data set in Section \ref{real_data}. Treatments 1 to 11 are defined in Figure \ref{fig:realdata}. Here the thickness of each edge $ab$ represents the associated weight, ${w}_{ab}$. The aggregate weights, as presented in the box, were calculated using the methods described in Section \ref{NMA_model}. The values are quoted to 3 decimal places.  }
    \label{fig:RealAgg}
\end{figure}

We now apply the random-walk approach to the data set described in Section \ref{real_data}. Following R\"ucker and Schwarzer (2014) \cite{Rucker:2014}, we choose a fixed-effect model ($\tau^2=0$). The edge weights in the aggregate network were obtained using the methods described in Section \ref{NMA_model} and are shown in Figure \ref{fig:RealAgg}. 

\subsection{Evidence flows}
First, we use the random-walk approach described in Section \ref{RW_NMA} to obtain the evidence flows for a certain comparison.  We focus on the comparison of treatments 1 (tricyclic or tetracyclic antidepressants) and 3 (psychotherapy + usual care). To this end, we define the transition matrix for a random walker on the \textit{aggregate network} (Figure \ref{fig:RealAgg}) starting at node 1 and ending at node 3. Using Equation (\ref{eq:Pxy-V}) we find

\begin{align}
\label{eq:P_real1}
\boldsymbol{T}^{\text{(1-3)}} =
\begin{blockarray}{cccccccccccc}
& 1 & 2 & 3 & 4 & 5 & 6 & 7 & 8 & 9 & 10 & 11 \\
\begin{block}{c(ccccccccccc)}
    1 & 0 &	0 &	0.298 &	0 &	0 &	0.173 &	0.070 &	0 &	0.368 &	0 &	0.091 \\
    2 & 0 &	0 &	0 &	0 &	0 &	0.517 &	0 &	0.335 &	0 &	0 &	0.149  \\
    3 & 0 &	0 &	1 &	0 &	0 &	0 &	0 &	0 &	0 &	0 &	0 \\
    4 & 0 &	0 &	0.563 &	0 &	0 &	0 &	0 &	0 &	0.437 &	0 &	0  \\
    5 & 0 &	0 &	0.232 &	0 &	0 &	0 &	0 &	0 &	0.768 &	0 &	0 \\
    6 & 0.039 &	0.321 &	0.046 &	0 &	0 &	0 &	0.014 &	0.210 &	0.262 &	0 &	0.107  \\
    7 & 0.162 &	0 &	0 &	0 &	0 &	0.147 &	0 &	0 &	0.156 &	0.535 &	0  \\
    8 & 0 &	0.498 &	0 &	0 &	0 &	0.502 &	0 &	0 &	0 &	0 &	0  \\
    9 & 0.060 &	0 &	0.562 &	0.068 &	0.109 &	0.190 &	0.011 &	0 &	0 &	0 &	0  \\
    10 & 0 &	0 &	0 &	0 &	0 &	0 &	1 &	0 &	0 &	0 &	0 \\
    11 & 0.093 &	0.420 &	0 &	0 &	0 &	0.487 &	0 &	0 &	0 &	0 &	0 \\
\end{block}
\end{blockarray}.
\end{align}
We have labelled the rows and columns according to the treatments they represent and we quote the values of the entries in the matrix to 3 decimal places. The third row of $\boldsymbol{T}^{\text{(1-3)}}$ is constructed such that once the walker reaches node 3 (the end node) it remains there indefinitely.

As described in Section \ref{RW_NMA}, the evidence flow through each direct comparison for the network comparison 1-3 is obtained from the expected net number of times a walker crosses each edge as it travels from node 1 to node 3 on the aggregate network (Figure \ref{fig:RealAgg}). The expected net number of times a walker crosses each edge can be estimated by simulating a large ensemble of random walkers, each moving independently as described by the transition matrix $\boldsymbol{T}^{\text{(1-3)}}$. For each walker we count the net number of times it crosses the designated edge, and we then subsequently average over all walkers. The more walkers we simulate, the more accurate our estimation. 

\begin{figure}
    \centering
    \includegraphics[width=0.75\linewidth]{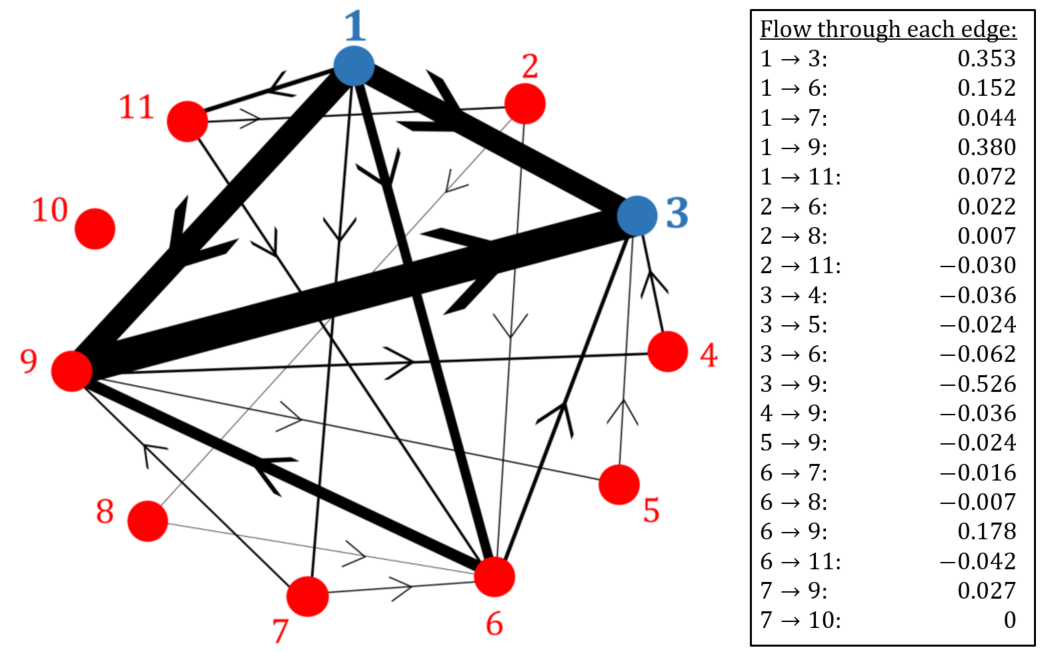}
    \caption{~The evidence flow network for the comparison of treatments 1 and 3 in the depression data set in Section \ref{real_data}. The thickness of each edge corresponds to the expected net number of times a random walker crosses each edge of the aggregate network in Figure \ref{fig:RealAgg} as it travels from node 1 to node 3. The direction of flow is indicated by the arrow. These values are summarised in the box and quoted to 3 decimal places. }
    \label{fig:RealFlow}
\end{figure}

Alternatively, we can use the analogy to electrical networks described in Section \ref{electric_RW}, to obtain an analytical result for this value in terms of electric current. These methods are described in more detail in Appendix \ref{App:CalcRWFlow}. We choose the analytical approach which results in the evidence flow network shown in Figure \ref{fig:RealFlow}.  We find that for the comparison of treatments 1 and 3, most of the evidence flows directly from 1 to 3 or indirectly via treatment 9. Comparing Figures \ref{fig:RealAgg} and \ref{fig:RealFlow} we observe that the pairwise comparison of treatments 7 and 10 is the only piece of direct evidence that has no influence on the network comparison 1-3.

The hat matrix of the aggregate model for this data is given in Appendix \ref{app:RealHat}. The flow network obtained from the row of the hat matrix corresponding to the comparison of treatments 1 and 3 is identical to the network in Figure \ref{fig:RealFlow}.

\subsection{Proportion contributions}
\label{apply-prop}
Next, we calculate the proportion contributions for the network comparison 1-3. To do this we first define the transition matrix for a random walker moving from node 1 to node 3 on the \textit{evidence flow network} (Figure \ref{fig:RealFlow}). From Equation (\ref{eq:U}) we find 
\setcounter{MaxMatrixCols}{11}

\begin{align}
\boldsymbol{U}^{\text{(1-3)}} = 
\begin{blockarray}{cccccccccccc}
& 1 & 2 & 3 & 4 & 5 & 6 & 7 & 8 & 9 & 10 & 11 \\
\begin{block}{c(ccccccccccc)}
    1 & 0 &        0 &  0.353 &         0 &         0 &  0.152 & 0.044 &         0 &  0.380  &        0 & 0.072\\
    2 & 0  &        0    &      0   &       0   &       0 &  0.755   &       0  & 0.245  &        0  &        0      &    0\\
    3 & 0  &         0   &       1   &       0   &       0    &      0   &       0  &        0   &       0   &       0  &        0\\
    4 & 0  &        0   &       1    &      0   &       0     &     0   &       0   &       0  &        0    &      0   &       0\\
    5 & 0   &       0  &        1   &       0    &      0    &      0    &      0    &      0   &       0     &    0     &     0\\
    6 & 0 &        0  & 0.259  &        0  &        0    &      0     &     0  &        0  & 0.741   &       0   &       0\\
    7 & 0   &       0    &      0    &      0    &      0  &  0.371    &      0   &       0 &  0.629    &      0     &     0\\
    8 & 0    &      0     &     0     &     0   &       0    &      1    &      0   &       0     &     0    &      0  &        0\\
    9 & 0   &       0   & 0.899 & 0.061 & 0.040   &       0    &      0    &      0   &       0    &      0      &    0\\
    10 & 0      &    0        &  0      &    0       &   0       &   0       &   0   &       0      &    0    &      1    &      0\\
    11 & 0  &  0.415     &     0        &  0      &    0  & 0.585    &      0     &     0      &    0     &     0     &     0 \\
\end{block}
\end{blockarray},
\end{align}
where we have again labelled the rows and columns. Matrix entries are quoted to 3 decimal places.  The third row indicates that once a walker reaches node 3 it remains there indefinitely.  Since treatment 10 is disconnected from all other nodes in the evidence flow network (Figure \ref{fig:RealFlow}), the probability of transitioning to this node from any other is zero. Similarly, if the walker starts at node 10, it remains there forever ($U_{\text{10-10}}^{\text{(1-3)}}=1$). 

The set of all possible paths that a random walker can take on the evidence flow network can be found using a recursive algorithm \cite{Andy:2011}. The probability with which the walker takes a particular path is calculated from Equation (\ref{eq:phi_RW}). This is the flow of evidence through that path. For the comparison of treatments 1 and 3 in the depression data set, we find 27 distinct paths. These paths and their associated flow $\phi_i$ make up the evidence streams presented in Table \ref{Tab:streams_average}.  We find $\phi_i \geq 0$ and $\sum_i \phi_i =1$. Using these values we can construct the network estimate $\hat{\theta}^{\mathrm{net}}_{\text{1-3}}$ as a linear combination of direct and indirect estimates following each evidence path listed in Table \ref{Tab:streams_average}. This leads to the same odds ratios as those quoted in R{\"u}cker and Schwarzer (2014) \cite{Rucker:2014} up to the precision provided.

\begin{table}[h]
\caption{~ Evidence streams (paths and their associated flow) for the network comparison of treatments 1 and 3 in the depression data set in Section \ref{real_data}. Results obtained from the random-walk (RW) approach are presented along with the results from three versions of the algorithm in Papakonstantinou et al \cite{Papakon:2018}. `Shortest' refers to the algorithm where paths are selected from shortest to longest. `Random' describes the variant in which paths are selected at random, and `Average' is the average over $10^8$ iterations of the Random algorithm. Values are rounded to 4 decimal places. The `Shortest' and `Random' algorithms fail to identify all possible paths, as indicated by the symbol `-'.}
\centering
\begin{tabular}{|l |l | c|c|c|c|}
\hline
\multirow{2}{*}{Stream, $S_i$}  &  \multirow{2}{*}{Path, $\pi_i$}  & \multicolumn{4}{c|}{Associated flow, $\phi_i$} \\ 
\cline{3-6} &  & RW approach & \multicolumn{3}{c|}{Algorithm}\\
\cline{4-6} & &(analytical) & Shortest & Random & Average \\
\hline
$S_1 = (\pi_1, \phi_1)$ & $1\xrightarrow{} 3$ & 0.3526 & 0.3526 & 0.3526 & 0.3526 \\
$S_2 = (\pi_2, \phi_2)$ &$1\xrightarrow{}6\xrightarrow{}3$& 0.0394 & 0.0622 & 0.0549 & 0.0303 \\
$S_3 = (\pi_3, \phi_3)$ &$1\xrightarrow{}6\xrightarrow{}9 \xrightarrow{} 3$& 0.1015 & 0.0901 & 0.0974 & 0.1091\\
$S_4 = (\pi_4, \phi_4)$ &$1\xrightarrow{}6\xrightarrow{}9 \xrightarrow{} 4 \xrightarrow{} 3$& 0.0069 & - & - & 0.0082\\
$S_5 = (\pi_5, \phi_5)$ &$1\xrightarrow{}6\xrightarrow{}9 \xrightarrow{} 5 \xrightarrow{} 3$& 0.0045 & - & - & 0.0048\\
$S_6 = (\pi_6, \phi_6)$ &$1\xrightarrow{}7\xrightarrow{}6 \xrightarrow{} 3$& 0.0042 & - & - & 0.0055\\
$S_7 = (\pi_7, \phi_7)$ &$1\xrightarrow{}7\xrightarrow{}6 \xrightarrow{} 9 \xrightarrow{} 3$& 0.0108 & - & - & 0.0061\\
$S_8 = (\pi_8, \phi_8)$ &$1\xrightarrow{}7\xrightarrow{}6 \xrightarrow{} 9 \xrightarrow{} 4 \xrightarrow{} 3$& 0.0007 & - & 0.0162 & 0.0024\\
$S_9 = (\pi_9, \phi_9)$ &$1\xrightarrow{}7\xrightarrow{}6 \xrightarrow{} 9 \xrightarrow{} 5 \xrightarrow{} 3$& 0.0005 & 0.0162 & - & 0.0021\\
$S_{10} = (\pi_{10}, \phi_{10})$ &$1\xrightarrow{}7\xrightarrow{}9 \xrightarrow{} 3$& 0.0246 & 0.0274 & 0.0274 & 0.0171\\
$S_{11} = (\pi_{11}, \phi_{11})$ &$1\xrightarrow{}7\xrightarrow{}9 \xrightarrow{} 4 \xrightarrow{} 3$& 0.0017 & - & - & 0.0060\\
$S_{12} = (\pi_{12}, \phi_{12})$ &$1\xrightarrow{}7\xrightarrow{}9 \xrightarrow{} 5 \xrightarrow{} 3$& 0.0011 & - & - & 0.0043\\
$S_{13} = (\pi_{13}, \phi_{13})$ &$1\xrightarrow{}9\xrightarrow{}3$& 0.3414 & 0.3798 & 0.3604 & 0.3656\\
$S_{14} = (\pi_{14}, \phi_{14})$ &$1\xrightarrow{}9\xrightarrow{}4 \xrightarrow{} 3$& 0.0231 & - & 0.0194 & 0.0090\\
$S_{15} = (\pi_{15}, \phi_{15})$ &$1\xrightarrow{}9\xrightarrow{}5 \xrightarrow{} 3$& 0.0153 & - & - & 0.0052\\
$S_{16} = (\pi_{16}, \phi_{16})$ &$1\xrightarrow{} 11\xrightarrow{} 2\xrightarrow{} 6\xrightarrow{} 3$ & 0.0058 & - & - & 0.0076 \\
$S_{17} = (\pi_{17}, \phi_{17})$ &$1\xrightarrow{} 11\xrightarrow{} 2\xrightarrow{} 6\xrightarrow{} 9 \xrightarrow{}3$ & 0.0150 & - & - & 0.0085\\
$S_{18} = (\pi_{18}, \phi_{18})$ &$1\xrightarrow{} 11\xrightarrow{} 2\xrightarrow{} 6\xrightarrow{} 9 \xrightarrow{}4 \xrightarrow{}3$ & 0.0010 & 0.0223 & - & 0.0034 \\
$S_{19} = (\pi_{19}, \phi_{19})$ &$1\xrightarrow{} 11\xrightarrow{} 2\xrightarrow{} 6\xrightarrow{} 9 \xrightarrow{}5 \xrightarrow{}3$ & 0.0007 & 0.0001 & 0.0225 &  0.0030\\
$S_{20} = (\pi_{20}, \phi_{20})$ &$1\xrightarrow{} 11\xrightarrow{} 2\xrightarrow{} 8\xrightarrow{} 6 \xrightarrow{}3$ & 0.0019 & - & 0.0073 & 0.0024\\
$S_{21} = (\pi_{21}, \phi_{21})$ &$1\xrightarrow{} 11\xrightarrow{} 2\xrightarrow{} 8\xrightarrow{} 6 \xrightarrow{}9 \xrightarrow{}3$ & 0.0049 & - & - & 0.0027\\
$S_{22} = (\pi_{22}, \phi_{22})$ &$1\xrightarrow{} 11\xrightarrow{} 2\xrightarrow{} 8\xrightarrow{} 6\xrightarrow{} 9\xrightarrow{} 4\xrightarrow{} 3$ & 0.0003 & - & - & 0.0012\\
$S_{23} = (\pi_{23}, \phi_{23})$ &$1\xrightarrow{} 11\xrightarrow{} 2\xrightarrow{} 8\xrightarrow{} 6\xrightarrow{} 9\xrightarrow{} 5\xrightarrow{} 3$ & 0.0002 & 0.0073 & - & 0.0010 \\
$S_{24} = (\pi_{24}, \phi_{24})$ &$1\xrightarrow{} 11\xrightarrow{} 6\xrightarrow{} 3$ & 0.0109 & - & - & 0.0163 \\
$S_{25} = (\pi_{25}, \phi_{25})$ &$1\xrightarrow{} 11\xrightarrow{} 6\xrightarrow{} 9 \xrightarrow{} 3$ & 0.0280 & 0.0288 & 0.0409 & 0.0170 \\
$S_{26} = (\pi_{26}, \phi_{26})$ &$1\xrightarrow{} 11\xrightarrow{} 6\xrightarrow{} 9 \xrightarrow{} 4 \xrightarrow{} 3$ & 0.0019 & 0.0132 & - & 0.0054 \\
$S_{27} = (\pi_{27}, \phi_{27})$ &$1\xrightarrow{} 11\xrightarrow{} 6\xrightarrow{} 9 \xrightarrow{} 5 \xrightarrow{} 3$ & 0.0013 & - & 0.0011 & 0.0033\\
\hline
\end{tabular}
\label{Tab:streams_average}
\end{table}

Table \ref{Tab:streams_average} also contains the streams identified by the algorithm in Papakonstantinou et al \cite{Papakon:2018} (see Section \ref{prop_1}). We present the results for three versions of the algorithm, Shortest, Random and Average. The results for the Random algorithm are obtained from one single run. Each result in the column labelled `Average' is an average over $10^8$ runs of the Random algorithm. From Table \ref{Tab:streams_average}, it is clear that the streams identified by the iterative algorithm depend on the order in which paths are selected. For this example, fewer than half of the possible paths are identified by the Shortest and Random algorithms (paths not detected are indicated by the symbol `-'). Therefore, these versions of the algorithm fail to take into account multiple evidence paths that contribute to the NMA (and potentially have a high risk of bias).  

Compared to the Shortest and Random versions of the algorithm, the Average algorithm produces results which are more similar to flows obtained from the random-walk approach. However, as described in Section \ref{prop_1}, the frequency with which a path is selected across different runs depends on whether it shares edges with other paths in the network. Therefore, the results of the Average algorithm do not necessarily converge to the results from the random-walk approach even as the number of iterations becomes large.

Using Table \ref{Tab:streams_average} and Equation (\ref{eq:PropCont}), we calculate the proportion contribution of each direct estimate to the network comparison of treatments 1 and 3 from the random-walk approach. These contributions are presented as percentages in the second column of Table \ref{Tab:prop_average}. The direct evidence from trials comparing treatments 1 and 3 has the largest contribution followed by indirect evidence from trials comparing 3 and 9, and 1 and 9. Table \ref{Tab:prop_average} also contains the proportion contributions obtained from the three versions of the algorithm (Shortest, Random and Average). As before, these results depend on the order in which paths are selected.

\begin{table}[h]
\caption{~ Proportion contributions, expressed as percentages, for the network comparison of treatments 1 and 3 in the depression data set. Results obtained from the random walk (RW) approach are presented along with the results from three versions of the algorithm in Papakonstantinou et al \cite{Papakon:2018}. Shortest refers to the algorithm where paths are selected from shortest to longest. Random is when paths are selected at random. Average is the average over $10^8$ iterations of the Random algorithm. Values are rounded to 1 decimal place.} 
\centering
\begin{tabular}{|c|c|c|c|c|}
\hline
\multirow{2}{*}{Direct evidence, $ab$}  & \multicolumn{4}{c|}{Proportion contribution, $p_{ab}^{(\text{1-3})}$} \\ 
\cline{2-5}   & RW approach & \multicolumn{3}{c|}{Algorithm}\\
\cline{3-5}  & & Shortest & Random & Average \\
\hline
1-3 & 35.3\% & 35.3\% & 35.3\% & 35.3\%\\
1-6 & 5.6 \% & 6.1 \% & 6.0\% & 5.5\%\\
1-7 & 1.3 \% & 1.2\% & 1.2\% & 1.3\%\\
1-9 & 18.4 \% & 19.0\%  & 18.7\% & 18.8\% \\
1-11 & 1.7\% & 1.5\% & 1.6\% & 1.7\% \\
2-6 & 0.5\% & 0.4\% & 0.4\% & 0.5\% \\
2-8 & 0.1\% & 0.1\% & 0.1\% & 0.1\% \\
2-11 & 0.6\% & 0.5\% & 0.5\% & 0.6\% \\
3-4 & 1.1\% & 0.6\% & 1.0\% & 0.9\% \\
3-5 & 0.7\% & 0.4\% & 0.4\% & 0.6\%\\
3-6 & 2.7\% & 3.1\% & 2.9\% & 2.5\%\\
3-9 & 22.6\% & 23.6\% & 23.2\% & 23.3\% \\
4-9 & 1.1\% & 0.6\% & 1.0\% & 0.9\% \\
5-9 & 0.7\% & 0.4\% & 0.4\% & 0.6\% \\
6-7 & 0.4\% & 0.3\% & 0.3\% & 0.4\% \\
6-8 & 0.1 \% & 0.1\% & 0.1\% & 0.1\% \\
6-9 & 5.1\% & 4.8\% & 5.0\% & 5.2\% \\
6-11 & 1.1\% & 1.0\% & 1.0\% & 1.1\% \\
7-9 & 0.9\% & 0.9\% & 0.9\% & 0.8\%\\
7-10 & 0\% & 0\% & 0\% & 0\%\\
\hline
\end{tabular}
\label{Tab:prop_average}
\end{table}

\section{Summary and Discussion} 
\label{discuss}

\subsection{The analogy between random walks and evidence flow, and the role of the graph theoretical model}

In this paper, we have presented a novel analogy between NMA and random walks. Edge weights from the aggregate graph theoretical NMA model define a transition matrix for a random walk on the network of evidence. The walker moves around on the aggregate network along edges corresponding to direct evidence. The movement of the random walker contains information about the propagation of evidence through the network. In particular, we have shown that the expected net number of times a walker crosses an edge can be interpreted as the evidence flow through the direct comparison represented by that edge. Therefore, we can obtain the elements of the hat matrix of the aggregate model from the random-walk process on the aggregate network.

The flow of evidence defined by K{\"o}nig et al (2013) \cite{konig:2013} is based on a two-step version of the standard frequentist NMA model (see Appendix \ref{App:Freq}). In the first step, the direct estimates are obtained by pooling evidence from trials making the same comparisons. For two-arm trials, a pairwise meta-analysis is performed. For multi-arm trials that compare a particular subset of treatments, an NMA is performed on the sub-graph described by the multi-arm trial design. The direct estimates are therefore separated into evidence that comes from two-arm trials and evidence from multi-arm trials. This is reflected in the hat matrix of this model. Consequently, in K{\"o}nig et al's evidence flow networks, the flow through multi-arm trials is displayed separately. This is an interesting feature but, as the authors note, it is only feasible for simple networks \cite{konig:2013}. 

In our definition of evidence flow, we have instead used a two-step version of the so-called \textit{graph theoretical} model \cite{Rucker:2012}. We make use of the fact that the adjusted weights describe a network of two-arm trials which is equivalent to the network of multi-arm trials. The direct estimates are then obtained from pairwise meta-analyses using the adjusted edge weights. The elements in the row of the hat matrix for a particular comparison then assign a single value of flow to each direct treatment comparison in the network. The flow through an edge therefore represents the combined contribution from all studies, two-arm and multi-arm, that make that comparison. While this means that the specific contribution of multi-arm studies is not displayed, our approach makes it easier to display evidence flow networks for graphs with a large number nodes, edges and multi-arm trials of varying designs. In addition, it is this property of the aggregate level graph theoretical approach that means we are able to make the analogy to random walks in the general case (i.e., networks including multi-armed trials).

As explained in Appendix \ref{App:Freq}, the standard NMA model, the graph theoretical model and the aggregate level versions of both these models, all yield the same network treatment effect estimates \cite{konig:2013, Rucker:2014}.  For networks containing exclusively two-arm trials, the hat matrices of the two aggregate level models are the same. Therefore, for these networks, the evidence flow networks we define are the same as those in K{\"o}nig et al.

The graph theoretical approach provides a straightforward visualisation of the flow of evidence for each treatment comparison. Random effects models and networks with multi-arm trials can be accounted for with no extra complications. For networks with both of these characteristics, heterogeneity needs to be combined with the original observed variances (i.e. one needs to use $\sigma_{i,ab}^2+\tau^2$ instead of $\sigma_{i,ab}^2$) before adjusting the weights to deal with multi-arm trials \cite{Rucker:2014, Gutman:2004}. 

\subsection{The random walk derivation of evidence streams overcomes the limitations of previous algorithms}

We have shown that the random-walk analogy for NMA leads to an analytical derivation of evidence streams. In doing so, we defined a second transition matrix, this time for a random walker moving on the evidence flow network. For each comparison of treatments $ab$ there is one separate evidence flow network. The network is directed and it has no cycles. Walkers can only move in one direction along each edge, according to the direction of flow. All paths on this graph start at $a$ and end at $b$. As the walker travels from $a$ to $b$ it moves along paths of direct and indirect evidence. Imagining a large number of independent random walkers undergoing this process, we interpret the proportion of walkers flowing through a particular path as the flow of evidence through that path, i.e., the flow of evidence through a path is the probability of a walker taking that path. This can be expressed analytically as the product of the transition probabilities along the edges that make up the path.

The analytical definition of evidence streams leads directly to an analytical derivation of the so-called proportion contributions defined in Papakonstantinou et al (2018)\cite{Papakon:2018}. The result is unambiguous in contrast with previously proposed algorithms whose output depends on the order in which paths are selected. Furthermore, individual runs of the algorithm in Papakonstantinou et al can fail to identify all paths of evidence on the evidence flow network. This means that in the calculation of proportion contributions, multiple paths of evidence and their potential bias are not taken into account. Running the algorithm many times and subsequently performing an average, we are eventually able to identify every path of evidence. However, the frequency with which a given path is selected depends strongly on the number of other paths with which it shares edges. As a result, the average flow obtained in this way does not accurately reflect the contribution of each path. The random-walk approach overcomes these limitations. All possible paths of evidence are identified and they are each assigned a value of flow that reflects the properties of the hat matrix. Therefore, all possible sources of bias are taken into account in the calculation of the proportion contributions. Since the result is purely analytical, the random-walk approach also offers superior computational efficiency.

For multi-arm trials, the method presented in Papakonstantinou et al (2018) na{\"i}vely treats each pairwise comparison in a multi-arm trial as an independent two-arm study \cite{Papakon:2018}. This does not account for correlations due to multi-arm trials. By instead using the adjusted weights from the graph theoretical model, we are able to define a network of two-arm trials that is equivalent to the original network of multi-arm trials. Therefore, an additional advantage of the methods presented in this paper, is that networks with multi-arm trials are handled more appropriately.

The CINeMA software currently relies on the algorithm in Papakonstantinou et al to calculate the relative contribution of studies with high, moderate and low risk of bias to each network treatment effect. Similarly, ROB-MEN (risk of bias due to missing evidence in network meta analysis\cite{Chiocchia:2021}) also uses the contribution matrix. Due to the significant advantages of the random-walk approach in deriving evidence streams we expect that applications such as these would benefit significantly in terms of accuracy and speed from the implementation of the method described in this paper. The recently updated PRISMA guidelines\cite{PRISMA2020} require systematic reviewers to assess their body of evidence for risk of bias. The results of our paper mean that existing 
software tools to help researchers make this assessment can now be made more reliable and efficient. We also plan to implement the aggregate hat matrix in \textit{netmeta}, along with the random-walk approaches to evidence flow and evidence streams.

\subsection{Potential future impact}

We believe that the analogy between NMA and random walks is interesting and that it provides new insight into NMA methodology. In our work we have explored the applications of only a small subset of the random-walk literature; there is, therefore, scope for the impact of this analogy to be investigated further. We hope that by presenting this analogy, more ideas will be shared between the two disciplines and additional practical applications of the random walk-approach will be developed in the future. 

For example, we have looked at the interpretation of the number of times a walker crosses each edge in the network. However, there is potentially also interest in investigating the number of times the walker visits each node. The random walk transition probabilities are proportional to the respective edge weights. Therefore, a walker is more likely to travel across an edge corresponding to a more precise treatment effect estimate. The expected number of times a walker visits a certain node will depend on how many connections the node has, and the weight (i.e., the inverse variance) associated with each of these connections. A node corresponding to a treatment that is involved in many direct comparisons will be visited more often than a node corresponding to a treatment with comparatively few connections. Furthermore, the larger the weight associated with the edges connected to a certain node, the more often the random walker will visit that node. Potentially, this value provides a measure of vertex centrality that accounts for both connectivity and the precision of treatment effect estimates. There may also be interest in measuring random walk variation. The variability in the paths traversed by a walker moving on the evidence flow network may indicate inconsistency between paths of indirect evidence.

In summary, by using the analogy to electrical networks as an intermediate step, we have made a novel connection between NMA and random walks. The interdisciplinary analogy provides new insight into NMA methodology. In particular, the analogy leads to an analytical derivation of the proportion contribution matrix without the ambiguity of existing numerical algorithms. Our approach can therefore be used to reliably quantify the contribution of individual study limitations to the resulting network treatment effects. We hope that this paper will provide a starting point for future developments of NMA methodology that can benefit from ideas in the random-walk literature.

\subsection*{Acknowledgements}

AD acknowledges funding by the Engineering and Physical Sciences Research Council (EPSRC UK), grant number EP/R513131/1. AN was supported by a Swiss National Science Foundation (SNSF) personal fellowship (P400PM\_186723). GR was supported by the German Research Foundation (DFG), grant RU 1747/1-2. TG is grateful for partial financial support by the Maria de Maeztu Program for Units of Excellence in R\&D (MDM-2017-0711). 

\subsection*{Data Availability Statement}
The data, results and associated codes used in this work can be found in the GitHub repository here \cite{Davies:Code}. For further details please contact the corresponding author.

\appendix

\section{Frequentist NMA}

\subsection{Standard and graph theoretical approaches (`reduce dimensions' vs. `reduce weights')}
\label{App:Freq}
\subsubsection{Standard frequentist NMA}
The standard frequentist approach to NMA is a regression analysis\cite{Lumley:2002, Salanti:2008, Efthimiou:2016}. The method relies on a design matrix $\boldsymbol{X}$ which is constructed to have full rank. Each $n_i$-arm trial contributes $n_i-1$ independent observations from which we aim to estimate $N-1$ independent network treatment effects. Therefore, the matrix $\boldsymbol{X}$ has dimensions  $\sum_i (n_i-1)\times(N-1)$. The `global baseline' treatment is chosen as treatment 1. Each column of $\boldsymbol{X}$ then refers to a treatment $ \in \{2,\hdots,N\}$. The rows represent the comparisons to the \textit{trial-specific} baseline in each study.  For a given row, the entry in the column corresponding to the treatment that is compared with the trial-specific baseline treatment is $+1$. If the trial specific baseline treatment is not the global baseline treatment, there is a $-1$ in the column corresponding to the trial-specific baseline. All other elements in the row are zero. 

The so-called `information matrix' is defined as $\boldsymbol{X}^\top \boldsymbol{V}^{-1} \boldsymbol{X}$ where $\boldsymbol{V}$ is the block-diagonal variance-covariance matrix. Each trial contributes an $(n_i-1)\times(n_i-1)$ block to $\boldsymbol{V}$ with observed variances on the diagonal and covariances (due to multi-arm trials)  off the diagonal. 

The hat matrix of the standard model is \cite{konig:2013, Rucker:2014}
\begin{align}
        \boldsymbol{H}^{\mathrm{(standard)}} = \boldsymbol{X}(\boldsymbol{X}^{\top} \boldsymbol{V}^{-1} \boldsymbol{X})^{-1} \boldsymbol{X}^{\top} \boldsymbol{V}^{-1}.
\end{align}

\subsubsection{Graph theoretical approach}

R{\"u}cker introduced an alternative \textit{graph theoretical} approach to NMA based on electrical network theory\cite{Rucker:2012}. This model is formulated around an edge-vertex incidence matrix $\boldsymbol{B}_0$ with dimensions $\sum_i \frac{n_i(n_i-1)}{2}\times N$, where $\sum_i \frac{n_i(n_i-1)}{2}$ is the total number of pairwise comparisons in the network. We write $\boldsymbol{B}_0$ for this matrix to distinguish it from the (similar) matrix $\boldsymbol{B}$ in the aggregate model described in Section~\ref{sec:aggregate} of the main paper. Each $n_i$-arm study contributes $\frac{n_i(n_i-1)}{2}$ rows to $\boldsymbol{B}_0$. Each column represents a treatment $\in \{1,\hdots,N\}$. Unlike the design matrix, $\boldsymbol{B}_0$ does not have full rank. Indeed, the elements in each row of $\boldsymbol{B}_0$ sum to zero \cite{Rucker:2014, Rucker:2012}. Entries of $\boldsymbol{B}_0$ are $+1$ in the column corresponding to the `baseline' treatment of the comparison represented by that row, and $-1$ in the column corresponding to the treatment compared to that baseline. 

We write $\boldsymbol{W}_0$ for the weight matrix of this model. Again, this is distinct from the matrix $\boldsymbol{W}$ in the main paper. 
$\boldsymbol{W}_0$ has dimensions $\left(\sum_i \frac{n_i(n_i-1)}{2}\right)\times\left(\sum_i \frac{n_i(n_i-1)}{2}\right)$ and contains on its diagonal the adjusted weights, $w_{i,ab}$, defined in the main paper. We obtain the adjusted weights from a method described in References \cite{Rucker:2012, Rucker:2014, Gutman:2004} which accounts for the correlations introduced by multi-arm trials. An important result of this method is that the adjusted weights describe the weights associated with a network of two-arm trials that is equivalent to the original network of multi-arm trials in the sense that the resulting relative treatment effect estimates from the network of two-arm trials are the same as those from the original network. By using these weights, we can therefore apply any NMA methodology that is only valid for networks of two-arm trials. The hat matrix of this model is,
\begin{align}
    \boldsymbol{H}^{\mathrm{(graph)}} = \boldsymbol{B}_0(\boldsymbol{B}_0^\top \boldsymbol{W}_0 \boldsymbol{B}_0)^{+} \boldsymbol{B}_0^\top \boldsymbol{W}_0.
\end{align}
\subsubsection{`Reduce dimension' vs `reduce weights'}
The design matrix $\boldsymbol{X}$ contains the same information about the structure of the network as $\boldsymbol{B}_0$ but has lower dimensions and full rank. For this reason R{\"u}cker and Schwarzer (2014)\cite{Rucker:2014} termed the standard model the `reduce dimension' approach. The alternative (graph theoretical) method relies on reducing the weights associated with observations from multi-arm trials. Therefore, this was termed the `reduce weights' approach \cite{Rucker:2014}. In R{\"u}cker and Schwarzer (2014) the authors proved that, although their respective hat matrices are different, the two approaches give rise to the same network treatment effect estimates and are, therefore, equivalent.

\subsection{Two-step models and evidence flow}
The concept of evidence flow was introduced by K{\"o}nig et al (2013) \cite{konig:2013}. Their approach was based on a two-step, or `aggregate', version of the \textit{reduce dimensions} (standard) model \cite{Lu:2011, Krahn:2013}:

{\it Step 1.} In the first step, evidence from all trials making the same comparisons is pooled. For two-arm trials, a pairwise meta-analysis is performed. For multi-arm trials with a particular design, an NMA is performed on the sub-graph described by the multi-arm design. The results from this first step define the direct evidence. 

{\it Step 2.} In step two, the direct estimates are used as observations in a linear regression model.

The hat matrix associated with this model defines the evidence flow. Since the direct evidence is separated into evidence from two-arm trials and evidence from multi-arm trials, K{\"o}nig et al display the flow through multi-arm trials separately on the evidence flow networks. The authors note that, with this approach, there is no unique way to represent evidence flow through multi-arm trials. Furthermore, explicitly showing multi-arm trials on evidence flow networks becomes increasingly difficult for large, highly connected networks.

In the main paper, we instead describe a two-step (aggregate) version of the \textit{reduce weights} (=graph theoretical) approach. The fact that the reduce weights model defines a matrix of two-arm trials that is equivalent to the matrix of multi-arm trials makes the two-step approach simpler. In the first step, we perform a pairwise meta-analysis across each edge using the adjusted weights. In the second step, we combine this aggregate (direct) data in a network meta-analysis. This approach yields exactly the same relative treatment effect estimates as the one-step reduce weights approach and, consequently, the reduce-dimensions approach. This equivalence also holds true for random effects models. One then needs to account for heterogeneity, i.e.  $\sigma_{i,ab}^2$ is replaced by $\sigma_{i,ab}^2+\tau^2$, before using the adjustment method\cite{Rucker:2012, Rucker:2014, Gutman:2004} to obtain the adjusted weights.  

For networks containing exclusively two-arm trials, the hat matrices from the two aggregate models are exactly equal. Therefore, in this scenario, our evidence flow networks are the same as those defined by K{\"o}nig et al \cite{konig:2013}. The differences arise in the presence of multi-arm trials. Our approach does not explicitly show the flow through multi-arm trials. Instead, the flow through each edge represents the pooled contribution from all studies that make that comparison. This is only made possible by using the reduce weights method to define a network of two-arm trials. Since each edge is associated with only one value of evidence flow, our approach makes it easier to construct evidence flow networks for complicated networks, i.e. those with many nodes, many connections, and many different multi-arm trials. This also makes it possible to calculate the proportion contribution matrix for networks of multi-arm trials. With the evidence flow networks defined by K{\"o}nig et al, this was not possible as the presence of multi-arm trials meant there were multiple values of flow associated with each edge.

\section{Heuristic argument for properties of the hat matrix and evidence flow}\label{sec:heuristic}

In this section we give a brief heuristic argument for the properties of the hat matrix in Section~\ref{flow}.  These properties were stated in Reference\cite{konig:2013}, an algebraic proof for some of the properties was given in Reference \cite{Papakon:2018}. We present our argument using the example in Figure~\ref{fig:PROP}, but this can be generalised to more complex networks.

\begin{figure}
    \centering
    \includegraphics[width=0.3\linewidth]{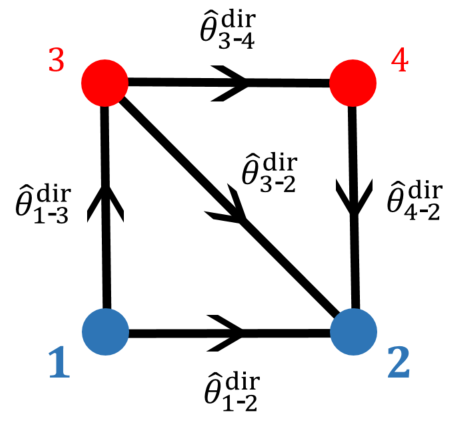}
    \caption{Meta analytic graph of the example in Figure~\ref{fig:PROP} (a). We focus on the comparison between treatments 1 and 2, as indicated by the blue colour of the nodes representing the treatments. Arrows show the sign conventions for the direction of evidence flow. Direct evidence for the relative treatment effects from the trial data are also indicated next to each comparison. 
    \label{fig:heuristic}}
\end{figure}

The network in Figure~\ref{fig:PROP} (a) is the evidence flow network for the comparison between treatments 1 and 2. It contains four nodes. For illustration and to fix sign conventions for the flow of evidence, the network is shown again in Figure~\ref{fig:heuristic}\hspace{-5pt}. Without loss of generality we assume that the direction of all edges are chosen such that $H_{cd}^{(1\text{-}2)}>0$ for all edges $cd$ shown in Figure~\ref{fig:heuristic}\hspace{-5pt}. This means that $f_{cd}^{(1\text{-}2)}=H_{cd}^{(1\text{-}2)}$ for all $cd$.

The three properties in Section~\ref{flow} translate into
\begin{enumerate}
    \item[1.] $f^{(1\text{-}2)}_{1\text{-}2}+f^{(1\text{-}2)}_{1\text{-}3}=1$;
    \item[2.] $f^{(1\text{-}2)}_{1\text{-}2}+f^{(1\text{-}2)}_{3\text{-}2}+f^{(1\text{-}2)}_{4\text{-}2}=1$;
     \item[3.] $f^{(1\text{-}2)}_{1\text{-}3}=f^{(1\text{-}2)}_{3\text{-}2}+f^{(1\text{-}2)}_{3\text{-}4}$ and $f^{(1\text{-}2)}_{3\text{-}4}=f^{(1\text{-}2)}_{4\text{-}2}$.
\end{enumerate}
We address these one-by-one. To do this we use Equation~(\ref{eq:linear_hat2}), $f_{cd}^{(1\text{-}2)}=H_{cd}^{(1\text{-}2)}$, and the above sign convention to note that

\begin{align}
\label{eq:linear_hat3}
    \hat{\theta}^{\mathrm{net}}_{\text{1-2}} = f_{\text{1-2}}^{(\text{1-2})}\hat{\theta}^{\mathrm{dir}}_{\text{1-2}} + f_{\text{1-3}}^{(\text{1-2})}\hat{\theta}^{\mathrm{dir}}_{\text{1-3}}  + f_{\text{3-2}}^{(\text{1-2})}\hat{\theta}^{\mathrm{dir}}_{\text{3-2}}  + f_{\text{4-2}}^{(\text{1-2})}\hat{\theta}^{\mathrm{dir}}_{\text{4-2}} + f_{\text{3-4}}^{(\text{1-2})}\hat{\theta}^{\mathrm{dir}}_{\text{3-4}}.    
\end{align}

\subsection{$f^{(1\text{-}2)}_{1\text{-}2}+f^{(1\text{-}2)}_{1\text{-}3}=1$}
Imagine we have one set of direct estimates,
\be
\boldsymbol{\hat\theta^{\text{dir}}}=(\hat{\theta}^{\mathrm{dir}}_{\text{1-2}},\hat{\theta}^{\mathrm{dir}}_{\text{1-3}},\hat{\theta}^{\mathrm{dir}}_{\text{3-2}},\hat{\theta}^{\mathrm{dir}}_{\text{3-4}},\hat{\theta}^{\mathrm{dir}}_{\text{4-2}}),
\ee
resulting in a network estimate $  \hat{\theta}^{\mathrm{net}}_{\text{1-2}}$ via Equation~(\ref{eq:linear_hat3}).

Imagine now a different set of direct estimates
\be
\boldsymbol{\hat\theta'^{\text{dir}}}=(\hat{\theta}'^{\mathrm{dir}}_{\text{1-2}},\hat{\theta}'^{\mathrm{dir}}_{\text{1-3}},\hat{\theta}'^{\mathrm{dir}}_{\text{3-2}},\hat{\theta}'^{\mathrm{dir}}_{\text{3-4}},\hat{\theta}'^{\mathrm{dir}}_{\text{4-2}}),
\ee
such that
\BE\label{eq:shift}
\hat{\theta}'^{\mathrm{dir}}_{\text{1-2}}&=&\hat{\theta}^{\mathrm{dir}}_{\text{1-2}}+\Delta, \nonumber \\
\hat{\theta}'^{\mathrm{dir}}_{\text{1-3}}&=&\hat{\theta}^{\mathrm{dir}}_{\text{1-3}}+\Delta, \nonumber \\
\hat{\theta}'^{\mathrm{dir}}_{\text{3-2}}&=&\hat{\theta}^{\mathrm{dir}}_{\text{3-2}}, \nonumber \\
\hat{\theta}'^{\mathrm{dir}}_{\text{3-4}}&=&\hat{\theta}^{\mathrm{dir}}_{\text{3-4}}, \nonumber \\
\hat{\theta}'^{\mathrm{dir}}_{\text{4-2}}&=&\hat{\theta}^{\mathrm{dir}}_{\text{4-2}}.
\EE
We write $\hat{\theta}'^{\mathrm{net}}_{\text{1-2}}$ for the network estimate from the dataset $\boldsymbol{\hat\theta'^{\text{dir}}}$.

Using the sign convention in which $\hat{\theta}^{\mathrm{dir}}_{cd}$ denotes the effect of treatment $d$ minus that of $c$, Equation~(\ref{eq:shift}) indicates that the direct effect of treatment 2 compared to treatment 1 in the dataset $\boldsymbol{\hat\theta'^{\text{dir}}}$ is $\Delta$ units greater than in dataset $\boldsymbol{\hat\theta^{\text{dir}}}$. Similarly, the relative effect of treatment 3 relative to treatment 1 is $\Delta$ units higher. Given that treatments 2 and 3 are the only ones treatment 1 is compared to directly in this network (see Figure~\ref{fig:heuristic}\hspace{-5pt}) we would then expect 
\be
\hat{\theta}'^{\mathrm{net}}_{\text{1-2}}=\hat{\theta}^{\mathrm{net}}_{\text{1-2}}+\Delta.
\ee
Using Equation~(\ref{eq:linear_hat3}) and its analogue for the dashed treatment effects, we find
\be
\hat{\theta}'^{\mathrm{net}}_{\text{1-2}}-\hat{\theta}^{\mathrm{net}}_{\text{1-2}}=\Delta\left(f^{(1\text{-}2)}_{1\text{-}2}+f^{(1\text{-}2)}_{1\text{-}3}\right),
\ee
and we therefore conclude
\be
f^{(1\text{-}2)}_{1\text{-}2}+f^{(1\text{-}2)}_{1\text{-}3}=1.
\ee

\subsection{$f^{(1\text{-}2)}_{1\text{-}2}+f^{(1\text{-}2)}_{3\text{-}2}+f^{(1\text{-}2)}_{4\text{-}2}=1$}
We again imagine a second set of data, now with
\BE\label{eq:shift2}
\hat{\theta}'^{\mathrm{dir}}_{\text{1-2}}&=&\hat{\theta}^{\mathrm{dir}}_{\text{1-2}}+\Delta, \nonumber \\
\hat{\theta}'^{\mathrm{dir}}_{\text{1-3}}&=&\hat{\theta}^{\mathrm{dir}}_{\text{1-3}}, \nonumber \\
\hat{\theta}'^{\mathrm{dir}}_{\text{3-2}}&=&\hat{\theta}^{\mathrm{dir}}_{\text{3-2}}+\Delta, \nonumber \\
\hat{\theta}'^{\mathrm{dir}}_{\text{3-4}}&=&\hat{\theta}^{\mathrm{dir}}_{\text{3-4}}, \nonumber \\
\hat{\theta}'^{\mathrm{dir}}_{\text{4-2}}&=&\hat{\theta}^{\mathrm{dir}}_{\text{4-2}}+\Delta.
\EE
This means that treatment $2$ is now consistently doing better by $\Delta$ units in relation to all treatments it is compared to directly in the network. The overall effect of this must be that
\be
\hat{\theta}'^{\mathrm{net}}_{\text{1-2}}=\hat{\theta}^{\mathrm{net}}_{\text{1-2}}+\Delta,
\ee
i.e., the effect of treatment 2 relative to that of treatment 1 is now $\Delta$ units greater. Using again Equation~(\ref{eq:linear_hat3}) for the data sets $\boldsymbol{\hat\theta^{\text{dir}}}$ and $\boldsymbol{\hat\theta'^{\text{dir}}}$ respectively, we now have
\be
\hat{\theta}'^{\mathrm{net}}_{\text{1-2}}-\hat{\theta}^{\mathrm{net}}_{\text{1-2}}=\Delta\left(f^{(1\text{-}2)}_{1\text{-}2}+f^{(1\text{-}2)}_{3\text{-}2}+f^{(1\text{-}2)}_{4\text{-}2}\right),
\ee
from Equation~(\ref{eq:shift2}). Therefore
\be
f^{(1\text{-}2)}_{1\text{-}2}+f^{(1\text{-}2)}_{3\text{-}2}+f^{(1\text{-}2)}_{4\text{-}2}=1.
\ee
\subsection{$f^{(1\text{-}2)}_{1\text{-}3}=f^{(1\text{-}2)}_{3\text{-}2}+f^{(1\text{-}2)}_{3\text{-}4}$ and $f^{(1\text{-}2)}_{3\text{-}4}=f^{(1\text{-}2)}_{4\text{-}2}$}

The first of these identities can be shown by looking at
\BE\label{eq:shift3}
\hat{\theta}'^{\mathrm{dir}}_{\text{1-2}}&=&\hat{\theta}^{\mathrm{dir}}_{\text{1-2}}, \nonumber \\
\hat{\theta}'^{\mathrm{dir}}_{\text{1-3}}&=&\hat{\theta}^{\mathrm{dir}}_{\text{1-3}}-\Delta, \nonumber \\
\hat{\theta}'^{\mathrm{dir}}_{\text{3-2}}&=&\hat{\theta}^{\mathrm{dir}}_{\text{3-2}}+\Delta, \nonumber \\
\hat{\theta}'^{\mathrm{dir}}_{\text{3-4}}&=&\hat{\theta}^{\mathrm{dir}}_{\text{3-4}}+\Delta, \nonumber \\
\hat{\theta}'^{\mathrm{dir}}_{\text{4-2}}&=&\hat{\theta}^{\mathrm{dir}}_{\text{4-2}},
\EE
and by realising that this means that treatment 3 now performs $\Delta$ units worse compared to all treatments it is directly compared to. This cannot affect the network estimate treatment effect of 2 compared to 1, i.e., we expect $\hat{\theta}'^{\mathrm{net}}_{\text{1-2}}=\hat{\theta}^{\mathrm{net}}_{\text{1-2}}$. This leads to $f^{(1\text{-}2)}_{1\text{-}3}=f^{(1\text{-}2)}_{3\text{-}2}+f^{(1\text{-}2)}_{3\text{-}4}$.

The identity $f^{(1\text{-}2)}_{3\text{-}4}=f^{(1\text{-}2)}_{4\text{-}2}$ can be demonstrated in a similar way .
\section{Electric current and evidence flow}
\label{App:Current_Flow}

In this section we demonstrate the relationship between electrical current and evidence flow. Consider an electrical network with $N$ nodes and $K$ edges. We define the vector of nodal or `external' currents as $\boldsymbol{J}=(J_1,J_2,\hdots,J_N)^\top$. These represent currents flowing between a node of the network and an external sink or source. Our sign convention is such that a positive entry $J_a>0$ indicates that a current goes into node $a$, whereas if $J_a<0$, a current goes out of node $a$. We write $\boldsymbol{I}=(I_1,I_2,\hdots,I_K)^\top$ for the currents in the edges $k=ab, k=1,2,\hdots,K$. A positive value of $I_{ab}$ indicates a flow of current from $a$ to $b$, and we set $I_{ba}=-I_{ab}$.

We define $\boldsymbol{\mathcal{V}}=\{\mathcal{V}_{ab}\}$ as the vector of voltages (potential differences) across the edges. That is, $\mathcal{V}_{ab}=v_a - v_b$ where $v_a$ and $v_b$ are the potentials at nodes $a$ and $b$ respectively. Ohm's law \cite{Ohm:1827} can then be written as
\begin{align}
\label{eq:igv}
    \boldsymbol{I} = \boldsymbol{C\mathcal{V}},
\end{align}
where  $\boldsymbol{C}$ is the $K\times K$ diagonal matrix of conductances (inverse resistances, $C_{ab}=(R_{ab})^{-1}$).  Using this and Kirchhoff's laws, R{\"u}cker \cite{Rucker:2012} demonstrated that $\boldsymbol{\mathcal{V}}$ can be written as
\begin{align}
    \boldsymbol{\mathcal{V}} = \boldsymbol{B} (\boldsymbol{B}^\top \boldsymbol{C} \boldsymbol{B})^{+} \boldsymbol{J},
\end{align}
where $\boldsymbol{B}$ is the edge-incidence matrix of the network defined in Section \ref{NMA_model}. Substituting this into Ohm's Law (Equation (\ref{eq:igv})) yields the edge currents,
\begin{align}
\label{eq:IJ}
    \boldsymbol{I} = \boldsymbol{C} \boldsymbol{B} \left(\boldsymbol{B}^\top \boldsymbol{C} \boldsymbol{B}\right)^{+} \boldsymbol{J}. 
\end{align}

To make the analogy to evidence flow, we consider an electrical network with a battery attached across the nodes corresponding to the treatment comparison we are interested in. For comparison $ab$ the external current at node $a$ is $J_a=+1$, at $b$ we have $J_b=-1$. The current $J_c$ at every other node $c\notin\{a,b\}$ is zero. 

We can do this in turn for each of the $K$ edges in the network. For convenience we label these $k=1,\dots, K$. We write $\boldsymbol{J}^{(k)}$ for the vector of nodal currents resulting in a situation where the battery is connected to the start and end points of edge $k$.

We then have  $K$ relations of the form in Equation (\ref{eq:IJ}),
\begin{align}
\label{eq:IJi}
    \boldsymbol{I}^{(k)} = \boldsymbol{C} \boldsymbol{B} \left(\boldsymbol{B}^\top \boldsymbol{C} \boldsymbol{B}\right)^{+} \boldsymbol{J}^{(k)}. 
\end{align}
We collect the internal currents $\boldsymbol{I}^{(k)}$ in a $K\times K$ matrix $\boldsymbol{\tilde{I}}=\begin{pmatrix}
     \boldsymbol{I}^{(1)} & \boldsymbol{I}^{(2)} & \hdots & \boldsymbol{I}^{(K)}
\end{pmatrix}$. Similarly, we define the $N\times K$ matrix $\boldsymbol{\tilde{J}}=\begin{pmatrix}
     \boldsymbol{J}^{(1)} & \boldsymbol{J}^{(2)} & \hdots & \boldsymbol{J}^{(K)}
\end{pmatrix}$. We then have
\begin{align}
\label{eq:Itilde}
     \boldsymbol{\tilde{I}} = \boldsymbol{C} \boldsymbol{B} \left(\boldsymbol{B}^\top \boldsymbol{C} \boldsymbol{B}\right)^{+} \boldsymbol{\tilde{J}}.
\end{align}

As an example, consider a simple network of three nodes 1,2,3 and where all  possible edges (1-2, 1-3, 2-3) are present. Let $k=1$ represent the edge 1-2, $k=2$ represent 1-3, and $k=3$ represent 2-3. The matrix of nodal currents is then
\begin{align}
    \boldsymbol{\tilde{J}} = \begin{pmatrix}
         1 & 1 & 0\\
         -1 & 0 & 1\\
         0 & -1 & -1
    \end{pmatrix}.
\end{align}
Each row of $\boldsymbol{J}$ represents a node, and each column represents a different placement of the battery. The first column corresponds to a battery attached across edge 1-2. Therefore, there is a $+1$ in the row corresponding to node 1, a $-1$ in the row corresponding to node 2 and a 0 for node 3. Similar reasoning is used to construct the other columns. 

From this construction, it is clear that the matrix of nodal currents for this setup is equal to the transpose of the edge incidence matrix,
\begin{align}
    \boldsymbol{\tilde{J}}=\boldsymbol{B}^\top.
\end{align}
We can write the resulting matrix of edge currents in terms of its composite elements,
\begin{align}
\label{eq:I_elements}
     \boldsymbol{\tilde{I}} = \begin{pmatrix}
        I_{\text{1-2}}^{(\text{1-2})} & I_{\text{1-2}}^{(\text{1-3})} & I_{\text{1-2}}^{(\text{2-3})} \\
        I_{\text{1-3}}^{(\text{1-2})} & I_{\text{1-3}}^{(\text{1-3})} & I_{\text{1-3}}^{(\text{2-3})}\\
        I_{\text{2-3}}^{(\text{1-2})} & I_{\text{2-3}}^{(\text{1-3})} & I_{\text{2-3}}^{(\text{2-3})}
    \end{pmatrix},
\end{align}
where $I_{cd}^{(ab)}$ is the current through edge $cd$ when a battery is attached across edge $ab$. 

In the evidence flow analogy, we interpret the flow of current $I_{cd}^{(ab)}$ as the flow of evidence through edge $cd$ for the network comparison $ab$. If the analogy holds (a proof follows below), we can write the elements of the hat matrix in terms of the edge currents. For the simple example above we have
\begin{align}
\label{eq:H_elements}
    \boldsymbol{H} = \begin{pmatrix}
        I_{\text{1-2}}^{(\text{1-2})} &  I_{\text{1-3}}^{(\text{1-2})} & I_{\text{2-3}}^{(\text{1-2})} \\
        I_{\text{1-2}}^{(\text{1-3})} & I_{\text{1-3}}^{(\text{1-3})} & I_{\text{2-3}}^{(\text{1-3})}\\
        I_{\text{1-2}}^{(\text{2-3})} & I_{\text{1-3}}^{(\text{2-3})} & I_{\text{2-3}}^{(\text{2-3})}
    \end{pmatrix}.
\end{align}
From Equations (\ref{eq:I_elements}) and (\ref{eq:H_elements}), it is clear that we need to prove that $\boldsymbol{\tilde{I}}^\top = \boldsymbol{H}$. 

We now do this for a general setup. Taking the transpose of Equation (\ref{eq:Itilde}), we find
\begin{align}
    \boldsymbol{\tilde{I}}^\top = \boldsymbol{\tilde{J}}^\top \left((\boldsymbol{B}^\top \boldsymbol{C} \boldsymbol{B})^{+}\right)^\top \boldsymbol{B}^\top \boldsymbol{C}^\top.
\end{align}
From the definition of the pseudo-inverse it is possible to show that $(\boldsymbol{A}^+)^\top = (\boldsymbol{A}^\top)^+$ for a general matrix $\boldsymbol{A}$ (see Reference \cite{Stoer:2002}). Using $\boldsymbol{\tilde{J}}=\boldsymbol{B}^\top$ and the fact that matrices $\boldsymbol{C}$ and $\boldsymbol{L}=\boldsymbol{B}^\top \boldsymbol{C} \boldsymbol{B}$ are symmetric ($\boldsymbol{C}^\top=\boldsymbol{C}$ and $\boldsymbol{L}^\top=\boldsymbol{L}$) we find
\begin{align}
\label{eq:Itilde-trans}
    \boldsymbol{\tilde{I}}^\top &= \boldsymbol{B} \left(\boldsymbol{B}^\top \boldsymbol{C} \boldsymbol{B}\right)^{+} \boldsymbol{B}^\top \boldsymbol{C}.
\end{align}
We now recall that the hat matrix of the aggregate model is (see Equation~(\ref{eq:HAT}) in the main paper)
\begin{align}
\label{eq:hat_again}
    \boldsymbol{H} = \boldsymbol{B}(\boldsymbol{B}^\top \boldsymbol{W} \boldsymbol{B})^+\boldsymbol{B}^\top \boldsymbol{W}.
\end{align}
The weight associated with each edge in the aggregate network ${w}_{ab}$ is given by the conductance (=inverse resistance) of that edge $C_{ab}=R_{ab}^{-1}$, see Section~\ref{NMA_electric} in the main paper. The matrices $\boldsymbol{W}$ and $\boldsymbol{C}$ contain these weights and conductances on their respective diagonals ($\boldsymbol{W}=\mathrm{diag}({w}_{ab})$ and $\boldsymbol{C}=\mathrm{diag}(C_{ab})$), and we therefore have
\begin{align}
\label{eq:C-V}
    \boldsymbol{C}=\boldsymbol{W}.
\end{align}
Substituting this into Equation (\ref{eq:Itilde-trans}), we find
\begin{align}
    \boldsymbol{\tilde{I}}^\top = \boldsymbol{H},
\end{align}
which is what we wanted to prove.

\section{Random walks and electric networks}
\label{App:RW_electric}

In this section, we demonstrate the relationship between electric current and random walks. This relationship is well known \cite{Doyle:2000}, and we include it here for completeness.

\subsection{Dirichlet problem for electric circuits}\label{sec:Dirichlet_el}
We start from Ohm's law. Rather than using matrix notation as in Appendix \ref{App:Current_Flow}, we  formulate Ohm's law for the current $I_{cd}$ in the edge $cd$,
\begin{align}
\label{eq:ohms}
    I_{cd} = C_{cd}(v_c-v_d),
\end{align}
 where $v_{c}$ and $v_d$ are the potentials at nodes $c$ and $d$ respectively. We have used the sign conventions of Reference\cite{Doyle:2000} to define the direction of current. As mentioned above we have $I_{cd}=-I_{dc}$. 

In this section we focus on the scenario where a unit current flows into node $a$ (from the exterior) and out of node $b$ (to the exterior). No flows between the network and the exterior are possible at any other nodes. To create such a situation we imagine a battery connected to nodes $a$ and $b$. The potential at $b$ is set to zero, and that at $a$ is $v_a=v_a^*$, with $v_a^*$ such that the external current into $a$ is equal to unity (the external current out of $b$ is then also equal to unity). The asterisk indicates the choice of $v_a$ resulting in a unit current into $a$. An illustration of this setup is shown in Figure \ref{fig:RW-electric} (b) in the main paper.

We use the superscript $(ab)$ to indicate a battery attached across $ab$ as described above, that is, we use $I_{cd}^{(ab)}$.  Kirchhoff's law states that the total current at any node $c\neq a, b$ is zero,
\begin{align}
\label{eq:Kirchhoff}
    \sum_{d} I_{cd}^{(ab)}=0  \text{ } \text{ } \text{ } \text{ } \forall c \neq a,b.
\end{align}
Substituting Equation (\ref{eq:ohms}) into Equation (\ref{eq:Kirchhoff}) and rearranging yields for $c\neq a,b$ 
\begin{align}
\label{eq:v_P}
    v_c = \sum_d \frac{C_{cd}}{\sum_{x}C_{cx}} v_d = \sum_d v_d T_{cd},
\end{align}
where we have used the definition of transition probabilities in Equation (\ref{eq:Pxy}) in the main paper. 

One can define a Laplacian matrix for this setup, $\boldsymbol{L}^{(ab)}=\boldsymbol{\mathds{1}}-\boldsymbol{T}^{(ab)}$, where $\boldsymbol{\mathds{1}}$ is the identity matrix\cite{Masuda:2017}. A twice continuously differentiable function $\boldsymbol{f}: c\mapsto f_c$ is then called harmonic if it satisfies the Laplace equation \cite{Axler:2001}, $\boldsymbol{L}^{(ab)} \boldsymbol{f}=\boldsymbol{0}$.

Equation (\ref{eq:v_P}) indicates that the function $c\mapsto v_c$ is harmonic at all points $c\neq a,b$. It also has boundary values at $a$ and $b$: $v_a=v_a^*$ is chosen such that the current going into node $a$ from the exterior is one, and we have $v_b=0$. This constitutes a Dirichlet problem\cite{Kakutani:1945}. The uniqueness principle for Dirichlet problems then implies that $v_c$ is uniquely determined for all $c$, given the boundary conditions at $a$ and $b$. For further details see Reference \cite{Doyle:2000}.

\subsection{Dirichlet problem for random walks}
We will now show that a quantity related to the expected net number of times a random walker visits a particular node $c$ while travelling from $a$ to $b$ fulfills the same Laplace equation, and shares the same boundary conditions as the electric potentials in Section~\ref{sec:Dirichlet_el}. The uniqueness of the solution of the Dirchlet problem then allows one to establish the analogy between electric networks and random walks. We now describe this in more detail.

We consider a walker starting at node $a$ and reaching absorption when it arrives at node $b$. We write $u_c$ for the expected number of times the walker visits node $c$ before reaching $b$ (with the convention that the final arrival at $b$ does not constitute a visit to $b$, i.e., we have $u_b=0$). The following relation then holds for $c \neq a, b$,
\begin{align}
\label{eq:ux}
    u_c = \sum_d u_d T_{dc}.
\end{align}
This equation can be understood as follows: In order to arrive at node $c$ the walker must previously visit a neighbouring node $d$. The quantity $u_d$ is the expected number of times this occurs. From such a node $d$ the walker must then transition to $c$ to contribute to $u_c$. This occurs with probability $T_{dc}$. Summing over all $d$ results in Equation (\ref{eq:ux}). 

Equation (\ref{eq:ux}) is of a similar form to Equation (\ref{eq:v_P}) in the electrical network. However $T_{dc}$ appears on the right-hand side of Equation~(\ref{eq:ux}), whereas one has $T_{cd}$ in Equation~(\ref{eq:v_P}). We therefore write $T_{dc}$ in terms of $T_{cd}$. Using Equation (\ref{eq:Pxy}), the definition $C_{ab}=R_{ab}^{-1}$, and the fact that  $C_{cd}=C_{dc}$, we find
\begin{align}
    T_{dc} = \frac{T_{cd} \sum_{x}C_{cx}}{\sum_{x}C_{dx}}.
\end{align}
Substituting this into Equation (\ref{eq:ux}) and re-arranging gives
\begin{align}
\label{eq:ux_R}
    \frac{u_c}{\sum_{x}C_{cx}} = \sum_d \frac{u_d}{\sum_{x}C_{dx}} T_{cd}.
\end{align}
Therefore, the object $c\mapsto u_c/(\sum_{x}C_{cx})$ is harmonic at all points $c \neq a, b$. Given that $u_b=0$, we have the boundary condition $u_b/(\sum_{x}C_{bx}) = 0$. We note that Equation (\ref{eq:ux_R}) and the boundary condition $u_b=0$ can be derived for any quantity $\boldsymbol{u}$ that is proportional to the number of visits at the different nodes. The Laplace equation and the boundary condition therefore only fix $u_c$ up to a factor. The uniqueness theorem for the Dirichlet problem also confirms that $u_c/(\sum_{x}C_{cx})$ is proportional to $v_c$ from Section~\ref{sec:Dirichlet_el} for all $c$. The constant of proportionality is fixed by the boundary condition for $u_a$.

We now show that the choice $u_a=(\sum_{x}C_{ax}) v_a^*$ (with $v_a^*$ as in Section~\ref{sec:Dirichlet_el}) is required if we want $u_c$ to be the expected number of times a walker starting at $a$ visits node $c$ before it reaches $b$. This choice implies
\begin{align}
\label{eq:v_u}
    v_c = \frac{u_c}{\sum_{x}C_{cx}}
\end{align}
for all $c\neq a,b$ by virtue of the uniqueness theorem, and using Equations (\ref{eq:v_P}) and (\ref{eq:ux_R}). In other words, $u_c/(\sum_{x}C_{cx})$ is then not only proportional to $v_c$, but identical to $v_c$ for all $c$.

We now prove that this is the appropriate choice. All we need to check is that the normalisation of the $u_c$ is consistent with the interpretation of $u_c$ as the number of times the walker visits node $c$. To do this we keep in mind that the walker starts at $a$ and finishes at $b$. Over the course of the walk returns to node $a$ are possible. The net number of times the walker leaves node $a$ however must be one, given that it starts at $a$ and ends at $b$ (this is the number of times the walker leaves $a$ minus the number of times it arrives at $a$, not counting the initial placement of the walker at $a$). If $u_c$ is the number of times a walker visits $c$ during the walk, then the expected net number of departures from node $c$ is given by $\sum_d (u_c T_{cd} - u_d T_{dc})$. Therefore we must have $\sum_c (u_a T_{ac} - u_c T_{ca})=1$. This condition is necessary for the correct normalisation of the $u_c$, and it is also sufficient to verify that the boundary condition $u_a=(\sum_{x}C_{ax}) v_a^*$ delivers this. This is what we will do next.

The boundary condition $u_a=(\sum_{x}C_{ax}) v_a^*$ leads to Equation~(\ref{eq:v_u}) as explained above. Substituting Equation (\ref{eq:v_u}) into Ohm's law (Equation (\ref{eq:ohms})), we find 
\begin{align}
    I_{cd}^{(ab)} &= C_{cd} \left( \frac{u_c}{\sum_{x}C_{cx}} - \frac{u_d}{\sum_{x}C_{dx}} \right) \\
    &= u_c \frac{C_{cd}}{\sum_{x}C_{cx}} - u_d \frac{C_{dc}}{\sum_{x}C_{dx}},
\end{align}
where, in the second step, we have used $C_{cd}=C_{dc}$. Finally, using Equation (\ref{eq:Pxy}), we find
\begin{align}
\label{eq:i_u}
    I_{cd}^{(ab)} = u_c T_{cd} - u_d T_{dc}.
\end{align}
The setup in Section~\ref{sec:Dirichlet_el} is such that the current into node $a$ (from the exterior) is equal to one. This means that the total current from node $a$ to all its neighbours in the network is also one, $\sum_c I_{ac}^{(ab)}=1$. We conclude that $\sum_c (u_a T_{ac} - u_c T_{ca})=1$, confirming the correct normalisation of the $u_c$.

In Appendix \ref{App:CalcRWFlow} we show how to obtain these edge currents analytically.

\section{Calculating the flow of evidence using the random walk approach}
\label{App:CalcRWFlow}
\subsection{Details of the calculation}
The interpretation of the flow of evidence as a random walk can be stated as follows: For the network comparison of treatments $a$ and $b$, the hat matrix element $H_{cd}^{(ab)}$ that defines the flow of evidence through the direct comparison $cd$ (via Equation~(\ref{eq:f}) in the main paper) is equal to the expected \textit{net} number of times a random walker, starting at $a$ on the aggregate NMA network and walking until it reaches $b$, moves along the edge from $c$ to $d$.

In Section \ref{RW_NMA} of the main paper we demonstrated how to construct a transition matrix for a random walker on the \textit{aggregate network}. For a particular comparison $ab$, we can use the transition matrix $\boldsymbol{T}^{(ab)}$ to simulate a large ensemble of independent random walkers on the aggregate network starting their journey at $a$ and stopping once they reach $b$. For each walker we count the number of times it moves across the different network edges in each direction. From this, we find the net number of times the walker move along a particular edge. By averaging these values over all of the simulated random walkers, we obtain an estimate of the evidence flow network for this comparison. The more walkers we simulate, the better our estimate of the evidence flow. 

By using the analogy between random walks and electrical networks, we can also obtain an analytical result for the evidence flow. To do so we make use of the equations in Appendix \ref{App:RW_electric}. First, we apply a 1 volt battery between nodes $a$ and $b$ so that the voltage at $a$ is $v_a=1$ and at $b$ is $v_b=0$. With these boundary conditions we then solve the simultaneous equations described by Equation (\ref{eq:v_P}),
\begin{align}
\label{eq:v_c}
    v_c = \sum_d v_d T_{cd},
\end{align}
to obtain the nodal voltages, $v_c$, for all nodes $c\neq a,b$. Using Ohm's law, we find the edge currents for the case of a 1 volt battery,
\begin{align}
\label{eq:I_v1}
    {I'}^{(ab)}_{cd} = C_{cd}(v_c - v_d) = w_{cd}(v_c - v_d),
\end{align}
these are indicated by ${I'}^{(ab)}_{cd}$ to distinguish them from the normalised currents $I^{(ab)}_{cd}$ in Section \ref{App:RW_electric}. In Equation (\ref{eq:I_v1}) we have used the fact that the conductance of edge $cd$ is equal to the aggregate weight associated with that edge, $C_{cd}=w_{cd}$. To make the analogy to evidence flow we require that the total external current flowing into node $a$ is 1. Therefore, to obtain the required currents we must normalise the currents ${I'}^{(ab)}_{cd}$ by dividing through by the total current flowing into $a$ when $v_a=1$, that is
\begin{align}
\label{eq:I_norm}
    I_{cd}^{(ab)} = \frac{{I'}^{(ab)}_{cd}}{\sum_x {I'}^{(ab)}_{ax}}.
\end{align}
As shown in Appendix \ref{App:RW_electric}, these currents are equal to the expected net number of times a random walker crosses each edge $cd$. Therefore, from Equation (\ref{eq:I_norm}) we obtain an analytical expression for the evidence flow network in terms of random walkers as follows:
\begin{align}
\label{eq:fN}
    H_{cd}^{(ab)}=\overline{N_{cd}^{(ab)}} =\frac{w_{cd} (v_c - v_d)}{\sum_{x} w_{ax} (v_a - v_x)},
\end{align} 
and $f_{cd}^{(ab)}$ is obtained from $H_{cd}^{(ab)}$ via Equation~(\ref{eq:f}). The potentials $v_x$ are obtained from Equation~(\ref{eq:v_c}).

\subsection{Implementing the calculation}
The above calculation can be written as a linear equation in matrix form. We provide this notation as it is useful for implementation. As above, we focus on the comparison $ab$ in a network of $N$ nodes such that our initial boundary conditions are $v_a=1$ and $v_b=0$. From Equation (\ref{eq:v_c}) we write, for $c\neq a, b$, 
\begin{align}
    v_c = \sum_d v_d T_{cd} = T_{ca} + \sum_{d \neq a, b} v_d T_{cd},
\end{align}
where we have inserted the known potentials, $v_a=1$ and $v_b=0$. Using the fact that $T_{cc}=0$ (for $c\neq b$) we eliminate the term $d=c$ on the right-hand side, and obtain  
\begin{align}
\label{eq:v_c_rearrange}
    v_c - \sum_{d \neq a,b,c} v_d T_{cd} = T_{ca}
\end{align}
for $c\neq a,b$. We collect the potentials $v_c$, $c\neq a,b$ in a vector $\boldsymbol{v}_{red}$ of length $N-2$. This is the vector of unknown potentials we wish to calculate. Similarly, we write the transition probabilities $T_{ca}$, $c\neq a,b$ as an $(N-2)$-vector, $\boldsymbol{T}^{(ab)}_{\cdot a}$. Therefore, we re-write Equation (\ref{eq:v_c_rearrange}) in matrix form as
\begin{align}
    (\boldsymbol{\mathds{1}} - \boldsymbol{T}^{(ab)}_{red})\boldsymbol{v}_{red} = \boldsymbol{T}^{(ab)}_{\cdot a}
\end{align}
where $\boldsymbol{\mathds{1}}$ is the $(N-2) \times (N-2)$ identity matrix, and $\boldsymbol{T}^{(ab)}_{red}$ is a reduced version of $\boldsymbol{T}^{(ab)}$ obtained from the full $N\times N$ transition matrix by removing the rows and columns corresponding to nodes $a$ and $b$. 

We then solve this equation for the vector of unknown potentials,
\begin{align}
    \boldsymbol{v}_{red} = (\boldsymbol{\mathds{1}} - \boldsymbol{T}^{(ab)}_{red})^{-1} \boldsymbol{T}^{(ab)}_{\cdot a}.
\end{align}
To obtain the full vector $\boldsymbol{v}$ of the potentials at all nodes, we use the fact that $v_a=1$ and $v_b=0$ and the entries of $\boldsymbol{v}_{red}$. 

The set of potential differences $v_c-v_d$ in Equation (\ref{eq:I_v1}) is then obtained by applying the edge-vertex incidence matrix to the vector of potentials, $\boldsymbol{B}\boldsymbol{v}$. Finally, multiplying by the weight matrix, $\boldsymbol{W}$, we obtain the vector of non-normalised edge currents (Equation (\ref{eq:I_v1}) in matrix notation),
\begin{align}
    {\boldsymbol{I}'}^{(ab)} = \boldsymbol{W}\boldsymbol{B}\boldsymbol{v}.
\end{align}
The normalised currents are then found by dividing through by the total current flowing from node $a$ into the network,
\begin{align}
    \boldsymbol{I}^{(ab)} = \frac{1}{\sum_x {I'}^{(ab)}_{ax}} {\boldsymbol{I}'}^{(ab)}.
\end{align}

\section{Application to real data}

\subsection{Evidence flow from hat matrix}
\label{app:RealHat}
The edge-vertex incidence matrix for the aggregate network of the depression data set (see Figure \ref{fig:RealAgg}) is
\begin{align}
\boldsymbol{B} = 
\begin{blockarray}{cccccccccccc}
& 1 & 2 & 3 & 4 & 5 & 6 & 7 & 8 & 9 & 10 & 11 \\
\begin{block}{c(ccccccccccc)}
    \text{1-3} & 1 &0 &	-1 &0 &	0 &	0 &	0 &	0 &	0 &	0 &	0 \\
    \text{1-6} & 1 &0 &	0 &	0 &	0 &	-1 &0 &	0 &	0 &	0 &	0\\
    \text{1-7} & 1 &0 &	0 &	0 &	0 &	0 &	-1 &0 &	0 &	0 &	0\\
    \text{1-9} & 1 &	0 &	0 &	0 &	0 &	0 &	0 &	0 &	-1 &0 &	0\\
    \text{1-11}& 1 &	0 &	0 &	0 &	0 &	0 &	0 &	0 &	0 &	0 &	-1\\
    \text{2-6}& 0 &	1 &	0 &	0 &	0 &	-1 &0 &	0 &	0 &	0 &	0\\
    \text{2-8} &0 &	1 &	0 &	0 &	0 &	0 &	0 &	-1 &0 &	0 &	0\\
    \text{2-11} &0 &	1 &	0 &	0 &	0 &	0 &	0 &	0 &	0 &	0 &	-1\\
    \text{3-4} &0 &	0 &	1 &	-1 &0 &	0 &	0 &	0 &	0 &	0 &	0\\
    \text{3-5}& 0 &	0 &	1 &	0 & -1 &0 &	0 &	0 &	0 &	0 &	0\\
    \text{3-6} &0 &	0 &	1 &	0 &	0 &	-1 &0 &	0 &	0 &	0 &	0\\
    \text{3-9}& 0 &	0 &	1 &	0 &	0 &	0 &	0 &	0 &	-1 &0 &	0\\
    \text{4-9}& 0 &	0 &	0 &	1 &	0 &	0 &	0 &	0 &	-1 &0 &	0\\
    \text{5-9}& 0 &	0 &	0 &	0 &	1 &	0 &	0 &	0 &	-1 &0 &	0\\
    \text{6-7}& 0 &	0 &	0 &	0 &	0 &	1 &	-1 &0 &	0 &	0 &	0\\
    \text{6-8}& 0 &	0 &	0 &	0 &	0 &	1 &	0 &	-1 &0 &	0 &	0\\
    \text{6-9}& 0 &	0 &	0 &	0 &	0 &	1 &	0 &	0 &	-1 &0 &	0\\
    \text{6-11}& 0 &	0 &	0 &	0 &	0 &	1 &	0 &	0 &	0 &	0 &	-1\\
    \text{7-9} &0 &	0 &	0 &	0 &	0 &	0 &	1 &	0 &	-1 &0 &	0\\
    \text{7-10}& 0 &	0 &	0 &	0 &	0 &	0 &	1 &	0 &	0 &	-1 &0\\
\end{block}
\end{blockarray}
\end{align}
where we have labelled the columns by the treatment and the rows by the direct treatment comparison that they represent. From the depression data (see Ref. \cite{Rucker:2014}) we obtain the adjusted weights using the adjustment method for multi arm trials (see Refs. \cite{Gutman:2004, Rucker:2012, Rucker:2014}). Using these weights, and Equations (\ref{eq:weight_mean}) and (\ref{eq:weight}) in the main paper, we perform a pairwise meta-analysis across each edge. The resulting aggregate weight matrix is \\

\scalebox{0.64}{%
$\boldsymbol{W} = 
\begin{blockarray}{ccccccccccccccccccccc}
& \text{1-3} &\text{1-6} & \text{1-7} & \text{1-9} & \text{1-11} & \text{2-6} &  \text{2-8} & \text{2-11} & \text{3-4} & \text{3-5} & \text{3-6} & \text{3-9} & \text{4-9}  & \text{5-9} & \text{6-7} & \text{6-8} & \text{6-9} & \text{6-11} & \text{7-9} & \text{7-10}\\
\begin{block}{c(cccccccccccccccccccc)}
\text{1-3}& 7.605& 0&       0&       0&       0&       0&       0&       0&       0&       0&       0&       0&       0&       0&       0&       0&       0&       0&       0&   0\\
\text{1-6} & 0&  4.432&       0&       0&       0&       0&       0&       0&       0&       0&       0&       0&       0&       0&       0&       0&       0&       0&       0&   0\\
\text{1-7} & 0 &      0& 1.785&       0&       0&       0&       0&       0&       0&       0&       0&       0&       0&       0&       0&       0&       0&       0&       0&   0\\
\text{1-9} & 0&       0&       0& 9.410&       0&       0&       0&       0&       0&       0&       0&       0&       0&       0&       0&       0&       0&       0&       0&   0\\
\text{1-11} & 0&       0&       0&       0& 2.322&       0&       0&       0&       0&       0&       0&       0&       0&       0&       0&       0&       0&       0&       0&   0\\
\text{2-6} &  0&       0&       0&       0&       0& 36.419&       0&       0&       0&       0&       0&       0&       0&       0&       0&       0&       0&       0&       0&   0\\
\text{2-8} & 0&       0&       0&       0&       0&       0& 23.576&       0&       0&       0&       0&       0&       0&       0&       0&       0&       0&       0&       0&   0\\
\text{2-11} & 0&       0&       0&       0&       0&       0&       0& 10.474&       0&       0&       0&       0&       0&       0&       0&       0&       0&       0&       0&   0\\
\text{3-4} & 0&       0&       0&       0&       0&       0&       0&       0& 13.559&       0&       0&       0&       0&       0&       0&       0&       0&       0&       0&   0\\
\text{3-5} & 0&       0&       0&       0&       0&       0&       0&       0&       0& 5.118&       0&       0&       0&       0&       0&       0&       0&       0&       0&   0\\
\text{3-6} & 0&       0&       0&       0&       0&       0&       0&       0&       0&       0& 5.187&       0&       0&       0&       0&       0&       0&       0&       0&   0\\
\text{3-9} & 0&       0&       0&       0&       0&       0&       0&       0&       0&       0&       0& 87.697&       0&       0&       0&       0&       0&       0&       0&   0\\
\text{4-9}  & 0&       0&       0&       0&       0&       0&       0&       0&       0&       0&       0&       0& 10.533&       0&       0&       0&       0&       0&       0&   0\\
\text{5-9} & 0&       0&       0&       0&       0&       0&       0&       0&       0&       0&       0&       0&       0& 16.946&       0&       0&       0&       0&       0&   0\\
\text{6-7} &0&       0&       0&       0&       0&       0&       0&       0&       0&       0&       0&       0&       0&       0& 1.620&       0&       0&       0&       0&   0\\
\text{6-8} & 0&       0&       0&       0&       0&       0&       0&       0&       0&       0&       0&       0&       0&       0&       0& 23.759&       0&       0&       0&   0\\
\text{6-9} & 0&       0&       0&       0&       0&       0&       0&       0&       0&       0&       0&       0&       0&       0&       0&       0& 29.714&       0&       0&   0\\
\text{6-11} & 0&       0&       0&       0&       0&       0&       0&       0&       0&       0&       0&       0&       0&       0&       0&       0&       0& 12.154&       0&   0\\
\text{7-9} & 0&       0&       0&       0&       0&       0&       0&       0&       0&       0&       0&       0&       0&       0&       0&       0&       0&       0& 1.713&   0\\
\text{7-10} & 0&       0&       0&       0&       0&       0&       0&       0&       0&       0&       0&       0&       0&       0&       0&       0&       0&       0&       0&   5.894\\
\end{block}
\end{blockarray}
$
}\\
 
\noindent where we have labelled the rows and columns by their respective direct treatment comparison. The values are rounded to 3 decimal places. \\

\noindent The hat matrix of the aggregate model is calculated using 
\begin{align}
    \boldsymbol{H} = \boldsymbol{B}(\boldsymbol{B}^\top \boldsymbol{W} \boldsymbol{B})^+\boldsymbol{B}^\top \boldsymbol{W}.
\end{align}
Therefore, for the depression data set we find \\

\scalebox{0.58}{%
$\boldsymbol{H} = 
\begin{blockarray}{ccccccccccccccccccccc}
& \text{1-3} &\text{1-6} & \text{1-7} & \text{1-9} & \text{1-11} & \text{2-6} &  \text{2-8} & \text{2-11} & \text{3-4} & \text{3-5} & \text{3-6} & \text{3-9} & \text{4-9}  & \text{5-9} & \text{6-7} & \text{6-8} & \text{6-9} & \text{6-11} & \text{7-9} & \text{7-10}\\
\begin{block}{c(cccccccccccccccccccc)}
    \text{1-3} & 0.353 &    0.152 &   0.044   &   0.380 &   0.072  &  0.022 &   0.007 &   -0.030 &   -0.036 &  -0.024 &  -0.062 &    -0.526 &  -0.036 &  -0.024 &  -0.016 &   -0.007 &     0.178 &   -0.042 &   0.027 & 0.000\\
    \text{1-6} & 0.261  &   0.236  &  0.052  &   0.339   &  0.111  &  0.035 &   0.011 &   -0.046 &    0.010 &   0.007  &  0.098 &     0.146 &   0.010 &  0.007 &  -0.040 &  -0.011 &   -0.514 &  -0.065 &   0.012 & 0.000\\
    \text{1-7} & 0.185  &   0.128 &    0.381  &   0.245 &   0.060  &  0.019 &  0.006 &  -0.025 &  0.0098 &  0.007 &   0.024 &    0.145 &  0.010 &  0.007  &   0.299 & -0.006  & -0.086  & -0.035  &  -0.321 & 0.000\\
    \text{1-9} & 0.307  &    0.160 &   0.046  &   0.412 &   0.075 &   0.024 &  0.008 &  -0.031  &    0.020 &   0.013 &  -0.022 &    0.296  &    0.020 &   0.013 &  -0.016 & -0.008  &   0.229 &  -0.044 &   0.030 & 0.000\\
    \text{1-11} & 0.235  &   0.213 &   0.046  &   0.305 &    0.201 &    -0.250 &  -0.081 &    0.331 &  0.009 &  0.006  &  0.089  &   0.132 &  0.009 &  0.006 &  -0.036  &  0.081 &   -0.463  &   0.468 &   0.011 & 0.000\\
    \text{2-6} &  0.005 &  0.004 & 0.001 &  0.006 &  -0.016 &    0.671 &    0.218  &   0.111 & 0.000 & 0.000 &  0.002 &  0.003 & 0.000 & 0.000 & -0.001  &  -0.218 & -0.009 &  -0.095 & 0.000 & 0.000\\
    \text{2-8} & 0.002 &  0.002 & 0.000 &  0.003  &  -0.008  &   0.337  &   0.607 &   0.056 & 0.000 &  0.000 &  0.001 &  0.001 & 0.000 &  0.000 & -0.000  &   0.393 & -0.005 &  -0.048 & 0.000    &     0.000\\
    \text{2-11} & -0.022 &  -0.020 & -0.004 &   -0.028 &   0.073 &    0.386 &    0.125 &    0.489 & -0.001 & -0.001 &  -0.008 &  -0.012 & -0.001 & -0.001 &  0.003    & -0.125 &   0.042 &    0.438 &   -0.001 & 0.000\\
    \text{3-4} & -0.020 &  0.003 &  0.001  &  0.014 &  0.002 & 0.000 & 0.000 & -0.001  &   0.587 &   0.016 &   0.017 &    0.359 &   -0.413 &   0.016 & 0.000 & -0.000  &  0.022 & -0.001 &  0.001     &    0.000\\
    \text{3-5} & -0.035 &  0.006 &  0.002  &  0.024 &  0.003 & 0.001 & 0.000 & -0.001  &  0.043 &     0.260 &   0.031  &   0.631 &   0.043  &   -0.740 & 0.000 & -0.000  &  0.039 & -0.002 &  0.002 & 0.000\\
    \text{3-6} & -0.091 &   0.084 &  0.008 &  -0.041 &   0.040 &   0.012 &  0.004 &  -0.016 &   0.045 &   0.030 &    0.161 &    0.673 &   0.045 &   0.030 &  -0.023 & -0.004 &   -0.692 &  -0.023 &  -0.015 & 0.000\\
    \text{3-9} & -0.046 &   0.007 &  0.003  &  0.032 &  0.003 &  0.001 & 0.000 & -0.001 &   0.056 &   0.037 &   0.039  &   0.822 &   0.056 &   0.037 & 0.000 & -0.000 &   0.051 & -0.002 &   0.003 & 0.000\\
     \text{4-9}  & -0.026 &  0.004 &  0.002  &  0.018 &  0.002 & 0.001  &  0.000 & -0.001 &   -0.532 &   0.021 &   0.022 &    0.463 &    0.468 &   0.021 & 0.000 &  -0.000 &   0.029 &  -0.001 &  0.002 & 0.000\\
    \text{5-9} & -0.012 &  0.002 & 0.001 &  0.007 & 0.001 & 0.000 & 0.000 & -0.000  &  0.013 &   -0.223 &  0.009 &    0.191 &   0.013 &    0.777 & 0.000 & 0.000 &   0.012 & -0.000 &  0.001   &   0.000\\
    \text{6-7} & -0.076 &   -0.108 &    0.329  &  -0.094 &  -0.051 &  -0.016 &  -0.005  &  0.021 & 0.000 & 0.000 &   -0.075 &  -0.001 & 0.000 & 0.000 &     0.338 &  0.005 &     0.428  &  0.030 &    -0.333 & 0.000\\
    \text{6-8} & -0.002 & -0.002 & -0.000 &  -0.003 &  0.008 &    -0.334  &   0.389 &  -0.055 & 0.000 & 0.000 & -0.001 & -0.001 & 0.000 & 0.000 &  0.000 &    0.611 &  0.005  &  0.047 & -0.000 &  0.000\\
    \text{6-9} & 0.046 &  -0.077 &  -0.005 &    0.072 &   -0.036 &   -0.011 &  -0.004 &    0.015 &   0.010 &   0.007 &   -0.121  &    0.150 &  0.010 &  0.007 &   0.023 &  0.004  &   0.743  &  0.021 &   0.018  &    0.000\\
    \text{6-11} & -0.026 &   -0.024 &  -0.005 &  -0.034 &   0.089 &   -0.285 &  -0.093  &   0.377 & -0.001 & -0.001 &  -0.010 &  -0.015 & -0.001 & -0.001 &  0.004  &  0.093 &   0.052 &    0.533 & -0.001   &      0.000\\
    \text{7-9} & 0.121  &  0.031 &   -0.334 &    0.166 &   0.015 &  0.005  & 0.002 & -0.006  &  0.010 &  0.007 &  -0.046 &    0.151 &    0.010 &  0.007  &  -0.315 &  -0.002 &    0.315 & -0.009 &    0.351 & 0.000\\
    \text{7-10} & 0.000 & 0.000 & 0.000 & 0.000 & 0.000 & 0.000 &  0.000 & 0.000 & 0.000 &  0.000 &  0.000 & 0.000 &   0.000 & 0.000 & 0.000 &  0.000 &      0.000    &     0.000  & 0.000 &    1.000\\
\end{block}
\end{blockarray}
$
}\\
\noindent The numerical values for the matrix entries are shown to 3 decimal places. The rows and columns are labelled by the treatment comparison they represent. The first row of the hat matrix refers to the network comparison of treatments 1 and 3. By comparing this row to Figure \ref{fig:RealFlow} in the main text, it is clear that the evidence flow network defined by the hat matrix is equivalent to the evidence flow network obtained from the random-walk approach.

\newpage

\bibliography{Ref_AMA.bib}%

\begin{thebibliography}{10}
\providecommand \doibase [0]{http://dx.doi.org/}%

\bibitem{TSD2}
Dias S, Welton NJ, Sutton AJ, Ades AE. {NICE DSU} technical support document 2:
  a generalised linear modelling framework for pairwise and network meta
  analysis of randomised controlled trials. Online;  2011.
\newblock Retrieved on September 2020 from \url{http://www.nicedsu.org.uk}.

\bibitem{DIAS:2018}
Dias S, Ades AE, Welton NJ, Jansen JP, Sutton AJ. {\it Network
  {M}eta-{A}nalysis for {D}ecision Making}.
\newblock Oxford: Wiley .
\newblock 2018.

\bibitem{SALANTI:2012}
Salanti G. Indirect and mixed treatment comparison, network, or multiple
  treatments meta analysis: many names, many benefits, many concerns for the
  next generation evidence synthesis tool. {\it {R}es {S}ynth {M}ethods}
  2012\string; 3\string: 80--97.

\bibitem{Lu:Ades:2004}
Lu G, Ades AE. Combination of direct and indirect evidence in mixed treatment
  comparisons. {\it {S}tat {M}ed} 2004\string; 23(20)\string: 3105--3124.

\bibitem{Hig:White:1996}
Higgins JPT, Whitehead A. Borrowing strength from external trials in a
  meta-analysis. {\it {S}tat {M}ed} 1996\string; 15(24)\string: 2733--2749.

\bibitem{Lumley:2002}
Lumley T. Network meta-analysis for indirect treatment comparisons. {\it Stat
  Med} 2002\string; 21\string: 2313--2324.

\bibitem{Rucker:2012}
R\"{u}cker G. Network meta-analysis, electrical networks and graph theory. {\it
  Res {S}ynth {M}ethods} 2012\string; 3\string: 312-324.

\bibitem{konig:2013}
König J, Krahn U, Binder H. Visualizing the flow of evidence in network
  meta-analysis and characterizing mixed treatment comparisons. {\it {S}tat
  {M}ed} 2013\string; 32(30)\string: 5414-5429.

\bibitem{Papakon:2018}
Papakonstantinou T, Nikolakopoulou A, Rücker G, et al. Estimating the
  contribution of studies in network meta-analysis: paths, flows and streams.
  {\it F1000Research} 2018\string; 7\string: 610.

\bibitem{Nikola:2020}
Nikolakopoulou A, Higgins JPT, Papakonstantinou T, et al. CINeMA: An approach
  for assessing confidence in the results of a network meta-analysis. {\it PLOS
  Med} 2020\string; 14(4)\string: 1-19.

\bibitem{Lu:2011}
Lu G, Welton NJ, Higgins J, White I, Ades A. Linear inference for mixed
  treatment comparison meta-analysis: a two-stage approach. {\it Res Synth
  Methods} 2011\string; 2(1)\string: 43-60.

\bibitem{Senn:2012}
Senn S, Gavini F, Magrez D, Scheen A. Issues in performing a network
  meta-analysis. {\it Stat Methods Med Res} 2012\string; 22(2)\string: 169-189.

\bibitem{Salanti:2008}
Salanti G, Higgins J, Ades A, Ioannidis J. Evaluation of networks of randomized
  trials. {\it Stat Methods Med Res} 2008\string; 17(3)\string: 279--301.

\bibitem{Salanti:2008b}
Salanti G, Kavvoura FK, Ioannidis JPA. Exploring the geometry of treatment
  networks. {\it {A}nn {I}ntern {M}ed} 2008\string; 148(7)\string: 544-553.

\bibitem{Davies:2020}
Davies AL, Galla T. Degree irregularity and rank probability bias in network
  meta-analysis. {\it Res Synth Methods} 2021\string; 12\string: 316– 332.

\bibitem{Tonin:2019}
Tonin FS, Borba HH, Mendes AM, Wiens A, Fernandez-Llimos F, Pontarolo R.
  Description of network meta-analysis geometry: A metrics design study. {\it
  PLoS ONE} 2019\string; 14(2)\string: 1-14.

\bibitem{Veroniki:2018}
Veroniki AA, Straus S, R{\"u}cker G, Tricco AC. Is providing uncertainty
  intervals in treatment ranking helpful in network meta-analysis?. {\it {J}
  {C}lin {E}pidemiol} 2018\string; 100\string: 122-129.

\bibitem{Kibret:2014}
Kibret T, Richer D, Bayene J. Bias in identification of the best treatment in a
  {B}ayesian network meta-analysis for binary outcome: a simulation study. {\it
  {C}lin {E}pidemiol} 2014\string; 6\string: 451--460.

\bibitem{Chiocchia:2021}
Chiocchia V, Nikolakopoulou A, Higgins JPT, et al. Tool to assess risk of bias
  due to missing evidence in network meta-analysis (ROB-MEN): elaboration and
  examples. {\it medRxiv} 2021\string: 1-34.

\bibitem{Mendes:2003}
Dorogovtsev S, Mendes J. {\it Evolution of Networks: From biological networks
  to the Internet and WWW}.
\newblock Oxford, UK: Oxford University Press .
\newblock 2003.

\bibitem{Newman:2018}
Newman M. {\it Networks}.
\newblock Oxford: Oxford University Press.
\newblock 2~ed. 2018.

\bibitem{Estrada:2011}
Estrada E. {\it The Structure of Complex Networks: Theory and Applications}.
\newblock Oxford, UK: Oxford University Press .
\newblock 2011.

\bibitem{Rucker:2015}
R{\"u}cker G, Schwarzer G. Ranking treatments in frequentist network
  meta-analysis works without resampling methods. {\it BMC Med Res Methodol}
  2015\string; 15(1)\string: 58.

\bibitem{Rucker:2020}
Rücker G, Petropoulou M, Schwarzer G. Network meta‐analysis of
  multicomponent interventions. {\it Biometrical J} 2020\string; 62\string:
  808-821.

\bibitem{netmeta:2021}
R{\"u}cker G, Krahn U, König J, Efthimiou O, Schwarzer G. {\it netmeta:
  Network meta-analysis using frequentist methods}. IMBI, University of
  Freiburg; Freiburg, Germany:  2021.
\newblock R package version 6.6-6, \url{https://github.com/guido-s/netmeta}.

\bibitem{Codling}
Codling EA, Plank MJ, Benhamou S. Random walk models in biology. {\it J R Soc
  Interface} 2008\string: 5813–834.

\bibitem{Okubo}
Okubo A, Levin SA. {\it Diffusion and Ecological Problems: Modern
  Perspectives}.
\newblock Springer, New York, NY, USA.
\newblock 2~ed. 2001.

\bibitem{Isichenko}
Isichenko MB. Percolation, statistical topography, and transport in random
  media. {\it Rev Mod Phys} 1992\string; 64\string: 961--1043.

\bibitem{Ewens}
Ewens WJ. {\it Mathematical Population Genetics I. Theoretical Introduction}.
\newblock New York, NY, USA: Springer .
\newblock 2010.

\bibitem{Mantegna}
Mantegna RN, Stanley HE. {\it An Introduction to Econophysics}.
\newblock Cambridge, UK: Cambridge University Press .
\newblock 1999.

\bibitem{Noh:2004}
Noh JD, Rieger H. Random Walks on Complex Networks. {\it Phys Rev Lett}
  2004\string; 92\string: 118701.

\bibitem{Lov:1994}
Lov{\'a}sz L. Random walks on graphs: A survey. Department of Computer Science,
  Yale University;  1994.
\newblock Available from
  \url{http://www.cs.yale.edu/publications/techreports/tr1029.pdf}.

\bibitem{MASUDA20171}
Masuda N, Porter MA, Lambiotte R. Random walks and diffusion on networks. {\it
  Physics Reports} 2017\string; 716-717\string: 1-58.
\newblock Random walks and diffusion on networks.

\bibitem{Kakutani:1945}
Kakutani S. Markov processes and the Dirichlet problem. {\it Proc Jap Acad}
  1945\string; 21\string: 227–233.

\bibitem{Kemeny:1966}
Kemeny J, Snell J, Knapp A. {\it Markov Chains}.
\newblock University Series in Higher Mathematics.New York: Van Nostrand .
\newblock 1966.

\bibitem{Kelly:1979}
Kelly F, Kelly F. {\it Reversibility and Stochastic Networks}.
\newblock Probability and Statistics Series.United Kingdom: J. Wiley .
\newblock 1979.

\bibitem{Doyle:2000}
Doyle PG, Snell L. Random walks and electric networks. arXiv:math/0001057;
  2000.

\bibitem{Linde:2013}
Linde K, Kriston L, R{\"u}cker G, Schneider A. Treatment of depressive
  disorders in primary care — a multiple treatment systematic review of
  randomized controlled trials. Bundesministerium für Bildung und Forschung
  (BMBF): Bonn Technical Report;  2013.
\newblock Retrieved on 12 April 2021 from
  \url{http://edok01.tib.uni-hannover.de/edoks/e01fb13/772211906.pdf}.

\bibitem{Rucker:2014}
R{\"u}cker G, Schwarzer G. Reduce dimension or reduce weights? Comparing two
  approaches to multi-arm studies in network meta-analysis. {\it Stat Med}
  2014\string; 33(25)\string: 4353-4369.

\bibitem{Efthimiou:2016}
Efthimiou O, Debray TP, {van Valkenhoef} G, et al. GetReal Methods Review
  Group. GetReal in network meta-analysis: a review of the methodology. {\it
  Res Synth Meth} 2016\string; 7(3)\string: 236-63.

\bibitem{Jackson:2012}
Jackson D, White IR, Riley RD. Quantifying the impact of between-study
  heterogeneity in multivariate meta-analyses. {\it Stat Med} 2012\string;
  31(29)\string: 3805–3820.

\bibitem{Gutman:2004}
Gutman I, Xiao W. Generalized inverse of the Laplacian matrix and some
  applications. {\it Bulletin T.CXXIX de l'Académie serbe des sciences et des
  arts} 2004\string; 29\string: 15-23.

\bibitem{Krahn:2013}
Krahn U, Binder H, K{\"o}nig J. A graphical tool for locating inconsistency in
  network meta-analyses. {\it BMC Med Res Methodol} 2013\string; 13(35)\string:
  1-18.

\bibitem{Urbano:2019}
Urbano M. {\it Introductory Electrical Engineering with Math Explained in
  Accessible Language.}ch.~19. Kirchhoff's Laws\string: 197-213; Hoboken, NJ,
  USA: John Wiley \& Sons, Ltd .
\newblock 2019.

\bibitem{Papakon:2020}
Papakonstantinou T, Nikolakopoulou A, Higgins JPT, Egger M, Salanti G. CINeMA:
  Software for semiautomated assessment of the confidence in the results of
  network meta-analysis. {\it Campbell Syst Rev} 2020\string; 16(1)\string:
  e1080.

\bibitem{Macfadyen:2005}
Macfadyen CA, Acuin JM, Gamble C. Topical antibiotics without steroids for
  chronically discharging ears with underlying eardrum perforations. {\it
  Cochrane DB Syst Rev} 2005\string; 4\string: 1-105.

\bibitem{Andy:2011}
{technical-recipes.com} . A recursive algorithm to find all paths between two
  given nodes in C++ and C. Online
  \url{https://www.technical-recipes.com/2011/a-recursive-algorithm-to-find-all-paths-between-two-given-nodes/};
  2011.

\bibitem{PRISMA2020}
Page MJ, McKenzie JE, Bossuyt PM, et al. The PRISMA 2020 statement: an updated
  guideline for reporting systematic reviews. {\it BMJ} 2021\string;
  372\string: 1-9.

\bibitem{Davies:Code}
Davies AL. Network meta-analysis and random walks: Data, codes and results.
  Online, GitHub;  2021.
\newblock Available from: \url{https://github.com/AnnieDavies/NMA_and_RW}.

\bibitem{Ohm:1827}
Ohm GS. {\it Die galvanische Kette, mathematisch bearbeitet}.
\newblock Berlin: T.H. Riemann .
\newblock 1827.

\bibitem{Stoer:2002}
Stoer J, Bulirsch R. {\it Introduction to Numerical Analysis}.
\newblock New York: Springer-Verlag.
\newblock 3~ed. 2002.

\bibitem{Masuda:2017}
Masuda N, Porter MA, Lambiotte R. Random walks and diffusion on networks. {\it
  Phys Rep} 2017\string; 716-717\string: 1-58.

\bibitem{Axler:2001}
Axler S, Bourdon P, Ramey W. {\it Harmonic Function Theory}.
\newblock New York: Springer .
\newblock 2001.

\end{thebibliography}

\end{document}